\documentclass[11pt]{article}
\usepackage[utf8]{inputenc}
\usepackage{amsfonts,epsfig}
\usepackage[hyphens]{url}
\RequirePackage{color}
\usepackage{enumitem}
\pdfoutput=1
\usepackage{hyperref}
\hypersetup{
	colorlinks=true,
	linkcolor=black,
	citecolor=black,
	filecolor=black,
	urlcolor=black,
}
\graphicspath{{fig/}}
\usepackage{relsize}
\usepackage{amssymb}
\usepackage{geometry}
\usepackage{sectsty}
\usepackage[font=it, labelfont=normal]{caption}
\usepackage{subcaption}
\usepackage{amsmath}
\DeclareMathOperator{\E}{\mathrm{E}}

\DeclareMathOperator{\R}{\mathbb{R}}

\DeclareMathOperator{\n}{\mathrm{normal}}
\usepackage[normalem]{ulem}
\sectionfont{\sffamily\upshape\large}
\subsectionfont{\sffamily\upshape\normalsize}
\subsubsectionfont{\sffamily\upshape\itshape\mdseries}
 \paragraphfont{\normalsize\sffamily\mdseries \itshape}
\makeatletter
\renewcommand\@seccntformat[1]{\csname the#1\endcsname.\quad}
\makeatother
 \usepackage[round]{natbib}
\usepackage{tikz}
\usetikzlibrary{bayesnet}
\usepackage[per-mode=symbol]{siunitx}

\makeatletter
\def\@maketitle{%
	\begin{center}%
		\let \footnote \thanks
		{\large \@title \par}%
		{\normalsize
			\begin{tabular}[t]{c}%
				\@author
			\end{tabular}\par}%
		{\small \@date}%
	\end{center}%
}
\makeatother

\title{\bf Making the most of imprecise  measurements:  Changing patterns of arsenic concentrations in shallow wells of Bangladesh from laboratory and field data \vspace{.1in}
}
\author{
	Yuling Yao\footnote{Department of Statistics, Columbia University.}  \and Rajib Mozumder\footnote{Lamont-Doherty Earth Observatory of Columbia University.}\and Benjamin  Bostick\footnotemark[3] \and Brian Mailloux\footnote{Environmental Sciences, Barnard College.} \and  Charles F.  Harvey\footnote{Department of Civil and Environmental Engineering, Massachusetts Institute of Technology.}  \and  Andrew Gelman\footnotemark[2]  \and Alexander van Geen\footnotemark[3]  
}
\date{\vspace{1em} 17 Jan 2021\vspace{-1em} }
\begin{document}\sloppy
	\maketitle
\thispagestyle{empty}

\begin{abstract}
 Millions of people in Bangladesh drink well water contaminated with arsenic. Despite the severity of this heath crisis, little is known about the extent to which groundwater arsenic concentrations change over time:  Are concentrations generally rising, or is arsenic being flushed out of aquifers? Are spatially patterns of high and low concentrations across wells homogenizing over time, or are these spatial gradients becoming more pronounced? 
To address these questions, we analyze a large set of arsenic concentrations that were sampled within a \SI{25}{\kilo\metre\squared} area of Bangladesh over time. 
We compare two blanket survey collected in 2000--2001 and 2012--2013 from the same villages but relying on a largely different set of wells.  The early set consists of 4574 accurate laboratory measurements, but the later set poses a challenge for analysis because it is composed of 8229 less accurate categorical measurements conducted in the field with a kit. 
We  construct a Bayesian model that jointly calibrates the measurement errors, applies spatial smoothing, and 
describes the spatiotemporal dynamic with a diffusion-like process model.
Our statistical analysis reveals that arsenic concentrations change over time and that their mean dropped from 110 to \SI{96}{\micro\gram\per\liter} over 12 years, although  one quarter of individual wells are inferred to see an increase.  
The largest decreases occurred at the wells with locally high concentrations where the estimated Laplacian indicated that the arsenic surface was strongly concave.  However, well with initially low concentrations were unlikely to be contaminated by nearby high concentration wells  over a decade.   We validate the model using a posterior predictive check on an external subset of laboratory measurements from the same 271 wells in the same study area available for 2000, 2014, and 2015. 

\end{abstract}

  \section{Introduction}
  Elevated levels of arsenic (As) in water pumped from shallow wells and consumed without treatment pose a serious threat to public health across many villages of South and Southeast Asia \citep{fendorf2010spatial, flanagan2012arsenic}. For lack of alternatives such as water treatment or a piped supply of safe water, rural residents of these regions with an unsafe well lowered their exposure to As mostly by switching their consumption to a neighbor's well that is low in As \citep{van2002promotion,chen2007reduction}. This is often a viable option because the sub-surface distribution of As in groundwater is highly heterogeneous. Well switching assumes, however, that wells are tested for As and that well users remember the test result, neither of which is necessarily the case. Another potential concern is that As levels in well water could potentially change over time. The few available studies relying on long-term monitoring indicate that groundwater As levels are by-and-large stable, but noteworthy exceptions have also been documented 
  \citep{dhar2008temporal,mcarthur2010migration,  mihajlov2020arsenic}. Groundwater flow patterns in shallow ($<$30 m deep) aquifers of the region have been highly perturbed and accelerated by irrigation pumping to grow rice during the winter season \citep{harvey2002arsenic, harvey2006groundwater}. It is therefore plausible to anticipate some changes over time, including the possibility of convergence of spatially highly variable As concentrations towards a local mean due to physical mixing and dispersion triggered by the daily turning on and off of numerous irrigation pumps within a given area.
  Such a convergence of As concentrations could pose a risk to residents relying on shallow wells that were low in As when they were tested and are rarely if ever tested again thereafter.     
  
\begin{figure} 
  	\centering
  	\includegraphics[width=\textwidth]{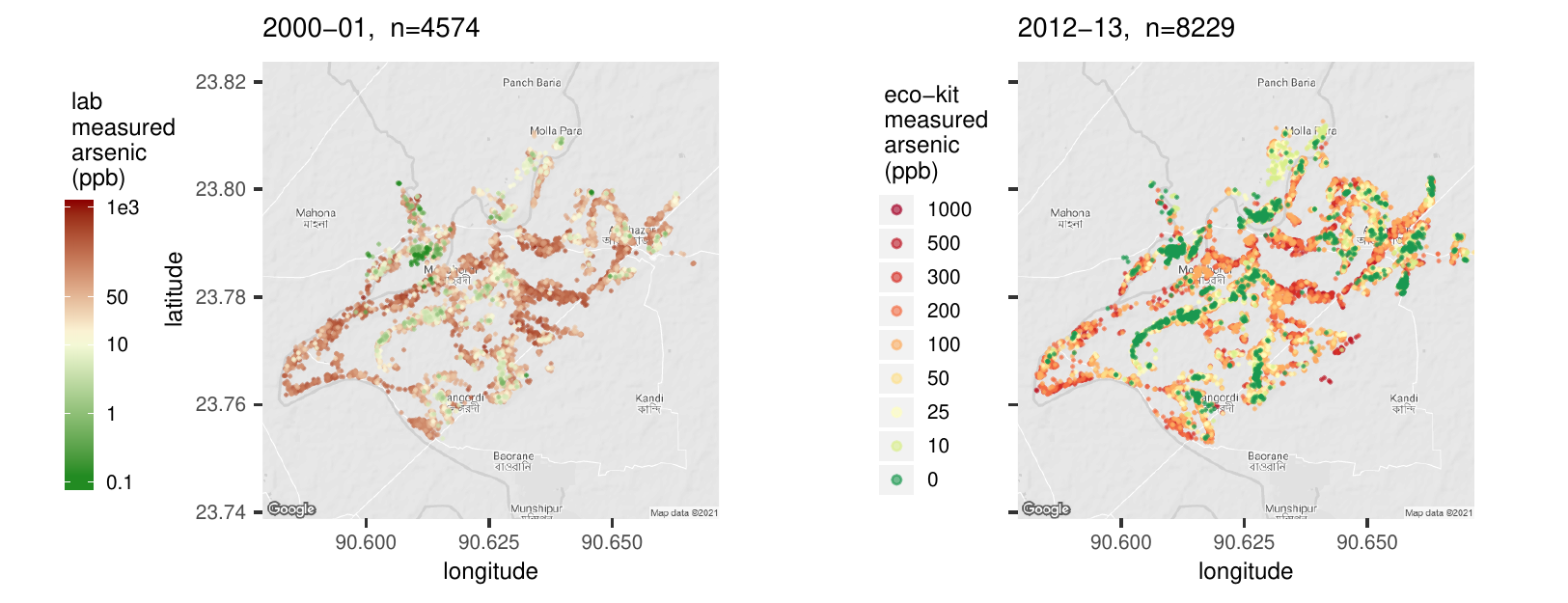}
  	\caption{ In 2000--2001,  a sample of size $n_1= 4574$ of wells  within a \SI{25}{\kilo\metre\squared} 
  		area of Bangladesh were analyzed by GFAA in laboratory.
  		The second survey took place in 2012--2013 where  $n_2=8229$ wells in the same area were measured by  Econo-Quick field kit  based on visual readings of a  test strip at 9 discrete levels (0, 10,   25,   50,  100,  200,  300,  500, 1000 ppb). Because they were measured  with different measurement precision  and at different sampling locations, these two datasets are not directly comparable.} \label{raw_spatial_reidentified_wells}
  \end{figure}

  To understand the evolution of spatial and temporal pattern of As in groundwater perturbed by irrigation pumping, we focus on a particularly well-studied \SI{25}{\kilo\metre\squared} area of Araihazar upazila (subdistrict) in Bangladesh. Groundwater samples from a total of 4827 shallow, mostly privately installed wells was sampled a first time in 2000--2001 and analyzed for As in the laboratory \citep{van2003spatial}. Most of these wells were replaced or re-installed by individual households in the subsequent decade \citep{van2014comparison}.
  In 2012--2013,  8228 shallow wells in the same area  were tested, but this time with the ITS Econo-Quick Arsenic field kit \citep{george2012evaluation} that is less precise and bins results in discrete categories (Figure \ref{raw_spatial_reidentified_wells}). A subset of wells were tested with the field kit and in the laboratory and can be used for intercalibration  \citep{Mozumder2019Impacts}. 
  Although the Econo-Quick field kit is cost-effective (about USD 0.30 per test) and sufficiently accurate for large-scale monitoring, their limited precision poses a challenge when trying to evaluate the stability of the distribution of As in shallow groundwater over time.  
  
  This paper seeks to quantify the changing pattern over a decade of As concentrations in a perturbed shallow aquifer from a combination of laboratory measurements and noisier test kits. The contribution of our work is threefold. First, we designed a flexible Bayesian model that incorporates a before-after comparison with different precision and measurement error.   Such modeling could be useful more generally when applied to field kits in environmental data analysis \citep[e.g.,][]{korfmacher2007reliability, landes2019field}.
  Second, we assessed the direction and magnitude of  groundwater mixing via a differential-equation-based dynamic.  The inference result suggests that wells with initially low As concentration are unlikely to be contaminated  by its  high As neighbors, which validates the current  recommendation for households with elevated arsenic levels to switch to a neighbor’s safe well.
  Third, as a byproduct of the Bayesian modeling, we calibrated  noisy individual  field kit test results communicated to households.

  Several projects have collected  As data  in Araihazar upazila,
  at different times and in different ways as follows. 
  \begin{enumerate}
  	\item \emph{{Blanket surveys}.} In 2000--2001,  a sample of size $n_1= 4574$ of groundwater from all wells in the study area was analyzed by graphite-furnace atomic absorption (GFAA) in a laboratory \citep{van2003spatial}. A subset of samples at the detection limit of \SI{5}{\micro\gram\per\liter} for As by this method was subsequently re-analyzed by inductively-coupled plasma mass spectrometry \citep[ICPMS, ][]{cheng2004rapid}. We denote the reading of the $i$-th well as $y_{1i}.$ In the second blanket survey conducted in 2012--2013, a group of trained workers tested $n_2=8229$ wells for As in the same area using the ITS Econo-Quick field kit. They recorded visual readings of a colored test strip relative to a reference scale showing 9 discrete levels (0, 10,   25,   50,  100,  200,  300,  500, \SI{1000}{\micro\gram\per\liter}). We denote the discrete field-kit measurement of well-$i$ as $w_{2i}.$ Most wells in the second survey were are different from wells sampled during the first study because of replacements and new installations over a decade \citep{van2014comparison}. The well index $i$ therefore does not corresponds the same unit in two surveys. The location of wells from both surveys was recorded with handheld GPS receivers with an accuracy of 10--50 m and is presented in a local Cartesian coordinate pair $(x^{\mathrm{E}}_{ti}, x^{\mathrm{N}}_{ti})$ for each sampled well. The depth of the well, as reported by households who paid for each section of pipe that went into the well's installation and therefore remember, is $d_{ti}$ in meters. 
  	\item  \emph{{Quality-control sample}.} The inaccuracy of field kits is not only attributable to  discretization of visual readings but reflects also potential groundwater matrix effects on the reading, the occasional wetting of the test strip, and differences in handling. For calibration, we saved  $n_{\mathrm{cal}}=944$  water samples from the second  blanket survey and analyzed them by ICPMS.  We denote the lab and kit values by $y^{\mathrm{cal}}_{i}$ and $w^{\mathrm{cal}}_{i}, ~1\leq i\leq n_{\mathrm{cal}}$ respectively. 
  	\item {\emph{Resampled  subset}.} A subset of 271 remaining wells from the 2000 blanket survey were re-identified in 2014 and 2015 with reasonable certainty based on the combination of a metal identification tag from 2000 that was still attached to the stem of the handpump along with a consistent location and installation year based on a conversation with the household that owns the well. These wells were therefore sampled a total of three times in 2000, 2014, and 2015 and are identified and labeled by consistent indices $i= 1,\dots, 271.$  The results from ICPMS analysis of all these samples, including re-analysis of the original set from 2000 for maximum consistency, are denoted by year $t$ and well $i$ as $\tilde y_{ti}$, for $t=2000,2014, 2015$.   The depth of the $i$-th well is  $\tilde d_{i}$ measured in 2000.
  	
  \end{enumerate}

  \begin{figure} 
  	\centering
  	\includegraphics[width=\textwidth]{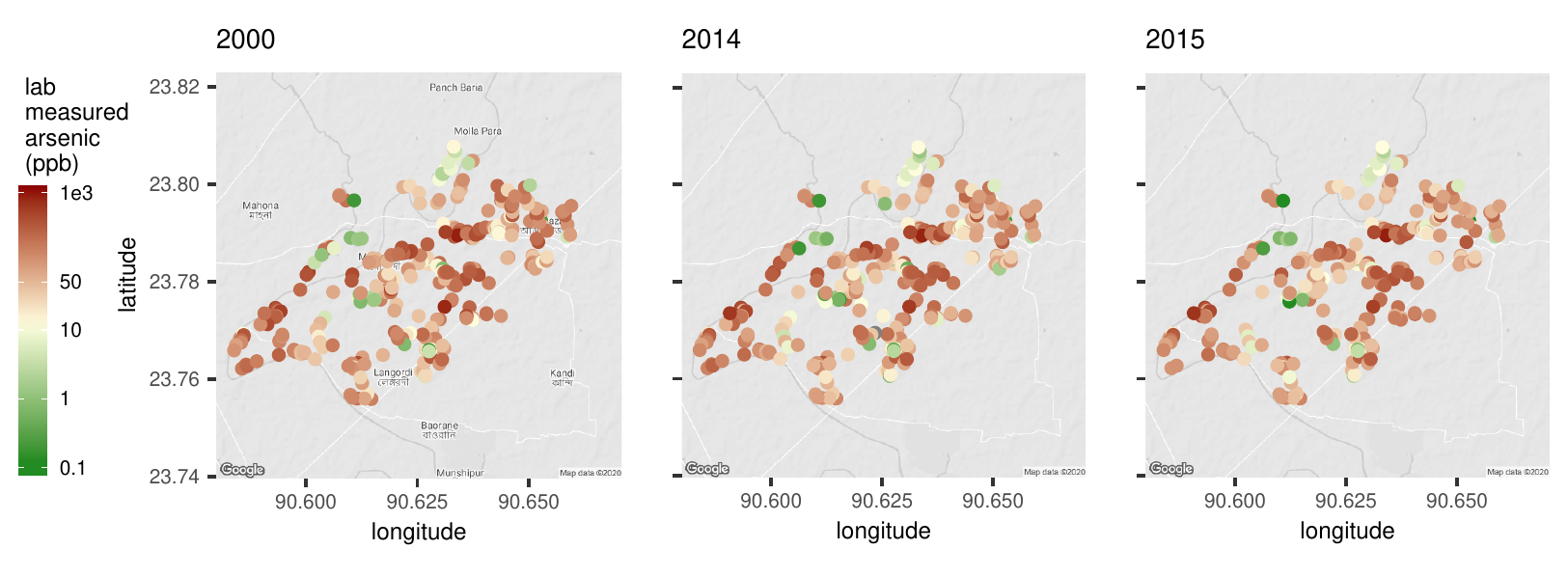}
  	\caption{308 wells in the 2000 blanket survey were re-identified in 2014 and measured in labs, among which 271 wells were tested again in 2015. Each dot in the graph represents one well.} \label{raw_spatial_reidentified_wells2}
  \end{figure}

  The remainder of this paper is organized as follows. In section \ref{sec_resample} we first explore the 271 resampled wells for which kit-measurement error is not a major issue. In section \ref{sec_error} we consider the measurement error models that calibrate the discrete field kit measurement. We propose a Bayesian hierarchical model in  section \ref{sec_reg} to infer spatiotemporal patterns from the noisy kit readings. Section \ref{sec_inf_result} presents the inference result of the blanket survey data, which reveals a general decline trend in As concentration across the study region. We conclude that physical homogenization is unlikely to be the driving force in this dynamic and hence do not need to be overly concerned about shallow low arsenic wells being affected by high As wells nearby. We discuss limitations and implications of our finding in section \ref{sec_discuss}.
  

  \section {Exploring the resampled   wells}\label{sec_resample}
  Because of the discreteness nature and large measurement error in the blanket survey, we first explore the  accurate lab measurements in the resampled 271 wells. Besides data exploration, we later reuse this dataset as an external validation. Figure \ref{raw_spatial_reidentified_wells2} visualizes the spatial distribution of arsenic values in the subset over time. All three maps show clusters of high, mixed, and low As concentrations that persisted over time.  The sample average As concentrations over time for the 271 resampled wells was \SI{100}{\micro\gram\per\liter} in 2000--2001 and \SI{90}{\micro\gram\per\liter} in 2014 and 2015. 

  \subsection{Regression to the mean vs. groundwater mixing}
\begin{figure}
  	\begin{center}
  		\includegraphics[width=\textwidth]{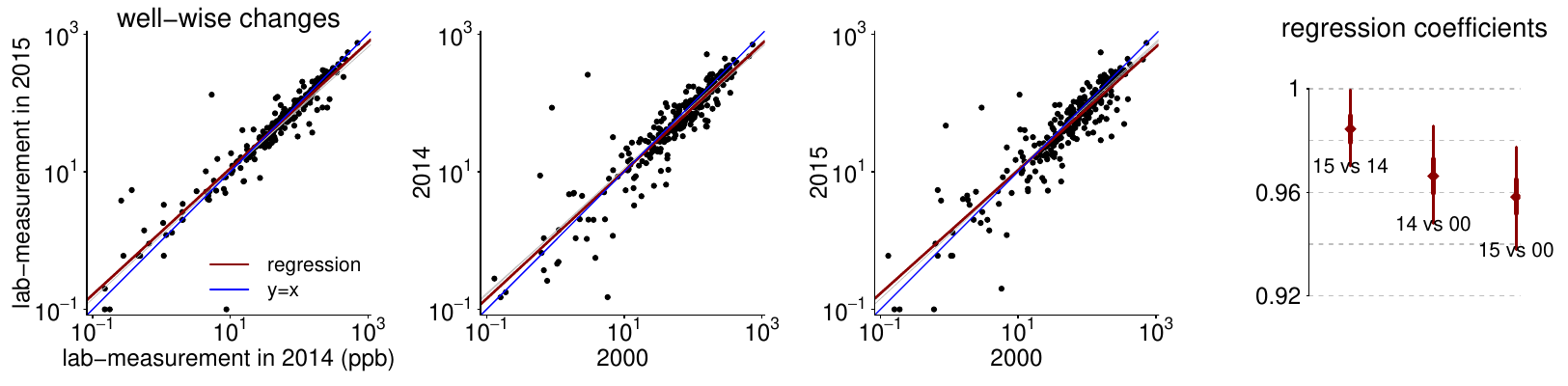}
  	\end{center}
  	\caption{``Regression to the mean." Lower values tend to increase and higher values tend to decrease. However, this does not separate the observational noise and the potential water mixing.} \label{temporal_change}
  \end{figure}

  The main public health concern  is whether physical water mixing could imperil wells that were initially safe when tested. To start, we run linear regressions on 3 pairs: 2014 versus 2000, 2015 versus 2014, and 2015 versus 2000 on the resampled dateset (Figure \ref{temporal_change}). The regression is in log scale to avoid the analysis driven by extremely large values.  
  The posterior means of the regression coefficients are 0.97, 0.98, and 0.96 for these three pairs, and their 95\% confidence intervals do not overlap with 1. The observed arsenic level in low-level wells tends to increase and in high-level wells tends to decrease in all three time periods. However, this pattern akin to regression to the mean could be the result of either measurement noise or natural fluctuations in each sampling period and is not necessarily an indication that physical mixing and homogenization is taking place.
  
  In Figure \ref{changes_over_time}, we compare  log As changes over time.  The well-wise changes in  2000--2014 are positively correlated with those in  2000--2015 but negatively correlated with 2014--2015.  This evidence suggests measurement error rather than physical mixing as the underlying cause of convergence of groundwater As concentrations. Both the short term and long term changes show large variations, with multiplicative shifts ranging from 20 to 1/20. 
  

  \subsection{Spatial smoothing using a Gaussian process and  autoregression using a spline}\label{sec_model_subset}
  To try to separate the pattern of water mixing, we  decompose the observed log arsenic values $\log \tilde y_{ti} ( t=2000, 2014, 2015, i=1, \dots, 271) $ into a spatial baseline arsenic level (the expected value conditional on the location), denoted by $\theta$,  the depth term, and residuals $\epsilon$.
  \begin{equation} \label{model_hmm}
  	\log \tilde y_{ti} = \theta_{t, i} +  \beta_{\mathrm{depth}} (\tilde d_i - d_0) +  \epsilon_{ti}, ~  \epsilon_{ti} \sim \n (0, \sigma_S).
  \end{equation}
  The spatial-dependent parameter $\theta_{t,i}$ is the baseline value in well $i$ and year $t$, which we will specify in the next paragraph. 
  The depth term describes the dependence of arsenic concentration on well depth via a linear form  $\beta_{\mathrm{depth}} (\tilde d_i - d_0)$, where $d_0=15.29$ meters is the pre-calculated mean depth of all wells. 
  The residual $\epsilon$ contains measurement errors, short-term fluctuations, or any other unobserved factors that can not be explained by the spatial distribution of wells. We model it by a normal distribution with standard deviation $\sigma_S$  (the \emph{short}-term).

  We model baseline arsenic concentration $\theta$ as a function of location $x \in \R^2$: the two dimensional location coordinate. That is 
  $\theta_{t,i}=\theta_{t}(x_i)$. We place a Gaussian process prior on  $\theta_{2000}(x)$, 
  with a squared exponential kernel, 
  \begin{equation} \label{model_gp}  
  	\theta_{2000}(x)  \sim \mathcal{GP} (\mu,  K ), ~~K(x_1, x_2 )= \alpha \exp\left( -\frac{ ||x_1-x_2||^2}{\rho^2}\right),
  \end{equation} 
  where $||x_1-x_2||^2$ is the Euclidean distance between two locations $x_1$ and $x_2$.  The Gaussian process is flexible to 
  characterize the spatial pattern.

  \begin{figure}
  	\begin{center}
  		\includegraphics[width=\textwidth]{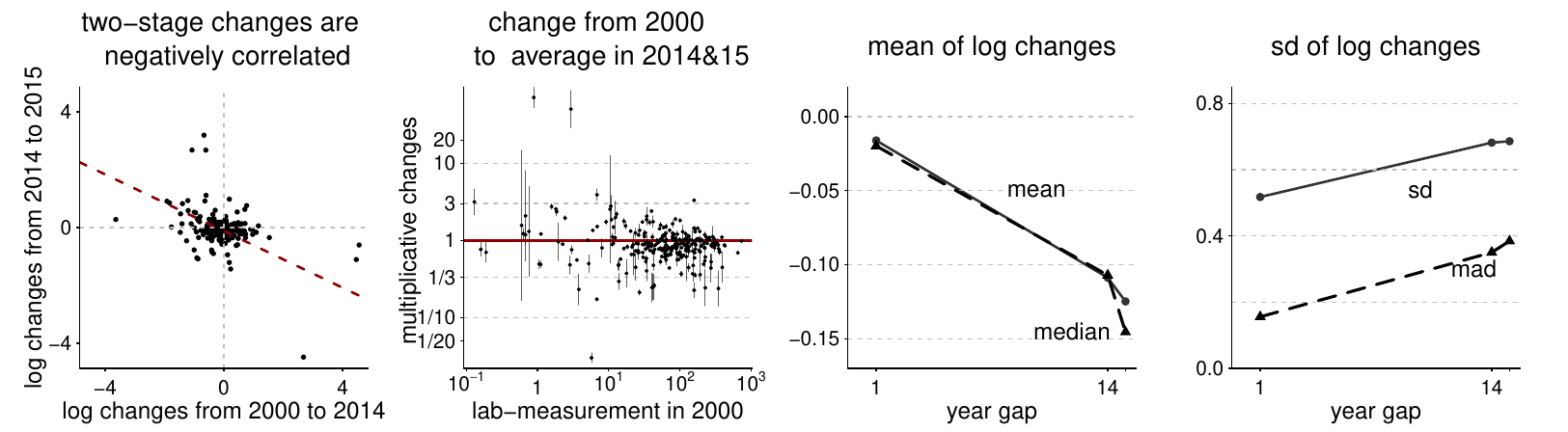}
  	\end{center}
  	\caption{ (1) The comparison of log change in 2000--2014 versus 2014--2015 per well. (2) The multiplicative change from 2000 to 2014 or 2015  per well as a function of the  the initial values. The two endpoints are the 2000--2014 and 2000--2015 changes and the dot is the average. (3--4):
  		Summary statistics (mean, median, standard deviation, median absolute deviation) of the well-level arsenic-changes as a function of year gaps. } \label{changes_over_time}
  \end{figure}

  From the summary statistics (mean, median, standard deviation, median absolute deviation) of the  well-level arsenic-changes (Figure \ref{changes_over_time}), the mean of the change from 2014 to 2015 (the left panel) is small compared to the change between 2000 and 2014. Therefore,  we assume 
  \begin{equation} \label{model_appro}  \theta_{2015, i}=\theta_{2014, i}, \quad i=1, \dots, 271.  \end{equation} 
  This approximation is computationally necessary because the problem is non-identified otherwise.
  
  Changes in well-water As concentrations driven by physical mixing would eventually lead to homogenization independently of observational noise. In order to approximate this mechanism, we consider an autoregression from $\theta_{2000}$ to $\theta_{2014}$. A strong negative  autocorrelation will support the water mixing hypothesis such that higher As values would  drop and lower As values would increase.
  
  In the exploration data analysis (Figure \ref{temporal_change}), we model a linear autoregression. To make it more flexible, we model the baseline change by a  cubic spline of the initial value:
  \begin{equation} \label{model_poly}  
  	\theta_{2014, i}-\theta_{2000, i}  =   \beta_0 \theta_{2000, i} +  \sum_{l=1}^L  \beta_l B_l (\theta_{2000, i}) + \mathrm{normal}(0, \sigma_{L}), \quad i=1, \dots, 271.
  \end{equation} 
  where $\{B_l\}_{l=1}^L$ is a collection of cubic spline basis functions.  Unlike typical spline smoothing directly applied to data, here we are fitting a spline regression on the latent variable $\theta$, whose range is unknown.  Nevertheless,  from the observational model \eqref{model_hmm},  it is reasonable to expect  that $\theta_{t,i}$ and  $\log \tilde y_{t,i}$ have a similar range.
  Hence we choose the internal knots of  the cubic spline to the $(0.1, 0.2,  \dots, 0.9)$ quantile of all observed $\log \tilde y_{t,i}$. We use a random-walk prior as on  spline coefficients $\beta_{l}$ to encourage smoothness, and place a weakly-informative prior on remaining parameters:
  \begin{align} \label{model_prior}  
  	\begin{gathered}
  		\beta_0, \beta_1 \sim \n (0,1), ~  \beta_{l+1}   \sim \n (\beta_l,0.5), \\
  		\sigma_S, \sigma_L \sim \mathrm{InvGamma}(3,3),   ~ \mu \sim \n (4, 1).
  	\end{gathered}
  \end{align}

  \begin{figure}
  	\centering 
  	\includegraphics[width=\textwidth]{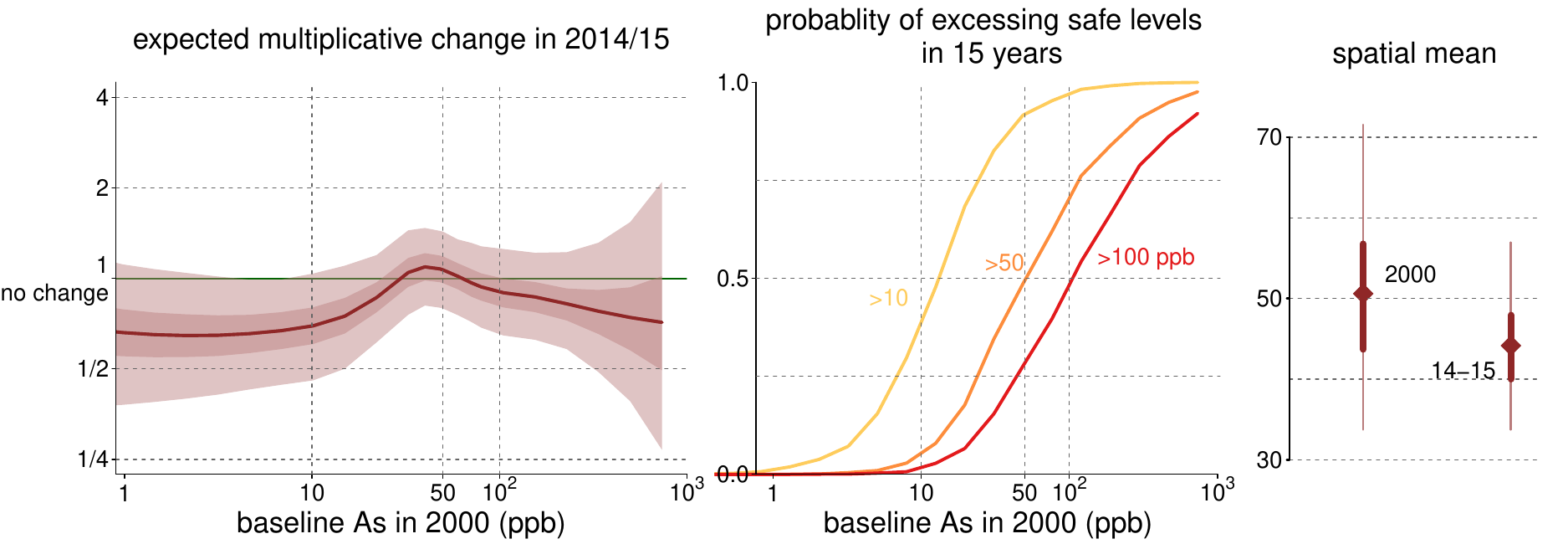}
  	\caption{ Left: posterior mean and intervals of the predicted change in 2014, given the well-arsenic level in 2000. Middle: the probability that a well would excess safety values in 15 years as a function of initial measurements. 
  		Right:  the global mean of $\theta_{2000, i}$ is  $\log 51$ and the mean of $\theta_{2014, i}$ is  $\log 44$. We label the y-axis by the usual linear scale (ppb) for readability. } \label{fig_spline}
  \end{figure}

  We sample from this joint posterior distribution in  
  Stan \citep{stan2020}, a platform for Bayesian modeling and computation. 

  \subsubsection*{Inference results for 271 resampled wells}
  

  The left panel of Figure \ref{changes_over_time} displays the posterior means, 50\%, and 95\%  confidence intervals of the autoregression cubic spline in \eqref{model_poly}, i.e., $\beta_0 \theta_{2000} +  \sum_{l=1}^L  \beta_l B_l (\theta_{2000})$  as a function of $\theta_{2000}$.
  It represents the expected value of log As changes in 2014 and 2015, given a well's initial
  baseline arsenic value in 2000. This result suggests that large initial  values were likely to drop over time. Perhaps more surprisingly, As concentrations at the low end of the range were more likely to decline as well, whereas intermediate values did not change appreciably. 

  Conditional on the the baseline As level in 2000, there is still a large amount of uncertainty of how the distribution of concentrations evolve due to both the long term regression noise $\sigma_L$ and the short term observational noise $\sigma_S$. 
  To help the interpretation of this uncertainty, we simulate in the middle panel a series of posterior draws to compute the probability of a realized laboratory test in a well  exceeding certain threshold in 15 years as a function of its initial baseline value: a well with initial value 10 ppb has a 40\% chance to exceed 10 ppb and 4\% chance to exceed 50 ppb in 2014's measurement. This graph provides an accessible way to communicate decision making:  despite all the temporal variation readily exists, a well is  unlikely to change from 10 ppb to 50 ppb in 14 years. In other words, a clearly safe/unsafe well (smaller than 10 or larger than  100 ppb) is nearly ensured to keep safe/unsafe and should be encouraged to switch to/away from. \citet{mailloux2020recommended} reach a similar conclusion using another approach as well as additional well-water arsenic data from different regions. 
  
  The right panel shows the fitted overall spatial mean of the latent baseline level $\theta$ in 2000 and 2014, with their 95\% confidence interval overlapped. This is in line with the 95\% confidence interval in the left panel showing an overall trend of mildly decreasing arsenic concentrations without ruling out that there was no change.

  The data also confirm the gradual increase in As concentrations with depth. The coefficient $\beta_\mathrm{depth}$ is estimated to be $0.03$ ($95$\% CI $(0.01,0.05)$) per meter. 
  
  The observational noise and suggestions of a ``regression to the mean'' do not exclude the possibility that the underlying As concentrations are truly converging due to physical water mixing. A limitation of the resampled wells is their small sample size. In the next section, we model the underlying stochastic process using two blanket surveys conducted in 2000--2001 and 2012--2013 that are much larger but partially measured by less accurate field kits.

  \section{A closer look at the two large surveys}
  
  \subsection{Kit calibration using the quality-control samples}\label{sec_error}
  The discrete kit readings $w_2$ in 2012--2013 are not directly comparable to the more precise laboratory measurements $y_1$ in the 2000--2001 survey.
  
\begin{figure}
  	\begin{center}
  		\includegraphics[width=\textwidth]{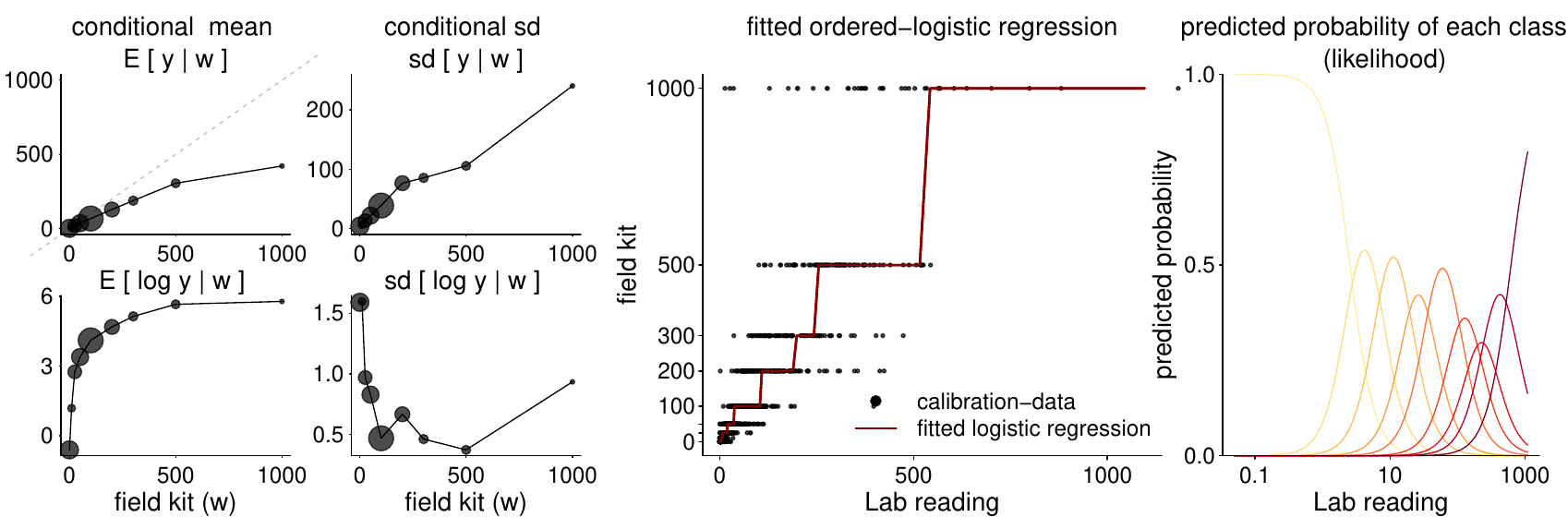} 
  	\end{center}
  	\caption{First two columns: conditional mean and standard deviation of lab measurement $y$ conditioning on the face value of eco-kit measurement $w$.   The dot size is the sample size in that category in the calibration set. Third column: the fitted ordered-logistic regression using the the quality-control sample.  The red line is the posterior prediction of $w$ given $y$, while the black dot is the observation in the calibration set.  Rightmost: the predicted probability of  each class  $\Pr(w=k|y, \hat \beta^\mathrm{cal}, \hat c^\mathrm{cal} )$, as a function of lab reading $y$.  Colors from shallow yellow to dark red represents kit-test classes from low to high.}\label{fig_cal}
  \end{figure}

  In the quality-control data, we observe both the kit measurement $w_{i}^{\mathrm {cal}}$ and  the lab measurement $y_{i}^{\mathrm {cal}}$ for $i=1,\dots, 944$.
  The  first two columns of Figure \ref{fig_cal} compares the face value of eco-kit and  laboratory measurements of the same water sample.  The kit reading exhibits a systemically negative bias and a considerable variance. For calibration, we treat the kit measurement $w$ as a nominal variable that only takes integer values $1,2,\dots,9$.
  We model the kit-measurement  $p(w_{i}^{\mathrm {cal}}| y_{i}^{\mathrm {cal}})$  by an ordered logistic regression with a flat prior and the following likelihood:
  \begin{align} \mathrm{Pr}(w_{i}^{\mathrm {cal}} \leq k\, |\, \beta^\mathrm{cal}, c^\mathrm{cal}, y_{i}^{\mathrm {cal}}  ) =\mathrm{logit}^{-1}\left( c^\mathrm{cal}_k + \beta^\mathrm{cal} \log y_{i}^{\mathrm {cal}}\right),\quad  k=1,2,\dots,9.
  \end{align} \label{eqn_ordered_log}
  The ordered logistic regression does not take the face value of the kit measurement as true value, but does keep the sequential order of the nominal readings. 
  The fitted result is shown in Figure \ref{fig_cal}.  With $n=944$ quality control samples, the joint posterior distribution of these 10 parameters is nearly a point mass.
  
  When it comes to the kit-measurement $w_{2i}$ in the large blanket survey that we will model in the next subsection, the lab measurement  is missing, and we will impute it probabilistically   from the ordered-logistic regression learned here.

  \subsection {Modeling the mixing process}\label{sec_reg} 
  We model the spatial arsenic distribution with a smoothing spline and superimpose a diffusion-like process to reflect changes in the distribution between 2000--2001 and 2012--2013. 
  
  \subsubsection*{Smoothing spline and its Laplacian}
  For the $i$-th well in the 2000--2001 survey at location $(x^{\mathrm{N}}_{1i}, x^{\mathrm{E}}_{1i})$, we decompose the log laboratory reading $\log y_{1i}$  into (a) the spatial baseline factor, modeled by a bivariate spline of ($x^{\mathrm{N}}_{1i}, x^{\mathrm{E}}_{1i}$),
  (b) the depth dependence, modeled by a linear function of depth $(d_{1i}-d_0)$, where $d_0$ is the average depth of all observed wells, and (c) an independent observational  noise.
  \begin{equation}\label{eq_log_y}
  	\log y_{1i}= \beta_{0}+ \sum_{l=1}^L\beta_{l} B_l(x^{\mathrm{N}}_{1i}, x^{\mathrm{E}}_{1i})+ 
  	\beta_\mathrm{depth} (d_{1i}-d_0)+
  	\n(0,\sigma_{\mathrm{obs}}),~1\leq i \leq n_1.
  \end{equation}

  We construct the bivariate spline from tensor products of one-dimensional cubic $B$-splines, with $L$ basis function $ B_l(x^{\mathrm{N}}_{2i}, x^{\mathrm{E}}_{2i})  =  B^\mathrm{N}_l(x^{\mathrm{N}}_{2i})  B^\mathrm{E}_l(x^{\mathrm{E}}_{2i})$.  After trimming, the number of product-basis-functions is $L=485$.  We detail the spline  knot choice in Appendix \ref{sec_spline}.

  From this spline model, for all sampling locations $x^{\mathrm{N}}_{2i}, x^{\mathrm{E}}_{2i}$ in the second survey, the counterfactual log baseline value is described by parameters $\theta_{1i}$: the expected log As value if well $i$ in the second had existed in 2000--2001 and we had tested its sample using lab reading.
  \begin{equation}\label{eq_bivariate_spline}
  	\theta_{1i}= \beta_{0}+ \sum_{l=1}^L\beta_{l} B_l(x^{\mathrm{N}}_{2i}, x^{\mathrm{E}}_{2i}), ~1\leq i \leq n_2.
  \end{equation}
  The spline model describes not only the value of  $\theta_{1i}$, but also the geometry of the static As value surface.  In particular,  the Laplace operator $\Delta= \frac{\partial^2}{\partial {(x^{\mathrm{N}})}^2} + \frac{\partial^2}{\partial {(x^{\mathrm{E}})}^2}$ describes the local curvature.   The Laplacian of basis functions comes in closed form  
  $  \Delta  B_l(x^{\mathrm{N}}_{2i}, x^{\mathrm{E}}_{2i})  = \Delta  B^\mathrm{N}_l(x^{\mathrm{N}}_{2i})  B^\mathrm{E}_l(x^{\mathrm{E}}_{2i})   + \Delta  B^\mathrm{E}_l(x^{\mathrm{E}}_{2i}) B^\mathrm{N}_l(x^{\mathrm{N}}_{2i}),$
  where  $\Delta  B^\mathrm{E}$  and $\Delta  B^\mathrm{N}$ is the second order derivative  of the one-dimensional cubic spline basis functions. Therefore, we  can extract pointwise Laplacian  $\delta_{i}$ as a linear transformation of spline coefficients: 
  \begin{equation}
  	\delta_{i}= \sum_{l=1}^L\beta_{l}   \Delta B_l(x^{\mathrm{N}}_{2i}, x^{\mathrm{E}}_{2i}), ~1\leq i \leq n_2,
  \end{equation}
  which reflects the As surface curvature in 2000--2001 at location $(x^{\mathrm{N}}_{2i}, x^{\mathrm{E}}_{2i})$.

  From the ordered logistic regression, we use a variable $\eta_{2i}$ to represent the log lab reading of well-$i$: the inferred log lab measurement if we had done the second survey in lab.  We plug-in the posterior mean of $\hat c^\mathrm{cal}_k, \hat \beta^\mathrm{cal}$ learned from the calibration  model \eqref{eqn_ordered_log},  and the likelihood of the missing lab measurement is  expressed by 
  \begin{equation}
  	\mathrm{logit(Pr}(w_{2i} \leq k  )) = \hat c^\mathrm{cal}_k + \hat \beta^\mathrm{cal}\eta_{2i},\quad  k=1,2,\dots,9, ~1\leq i \leq n_2.
  \end{equation}
  Like the observed  $\log y_{1i}$ in data model  \eqref{eq_log_y},  the hypothetical log lab reading $\eta_{2i}$ also contains  the noise-free baseline As surface $\theta_{2i}$, the depth dependence, and another independent Gaussian noise.
  \begin{equation}\label{eq_m21}
  	\eta_{2i} = \theta_{2i}+ \beta_\mathrm{depth} (d_{2i}-d_0)+   \n(0,\sigma_{\mathrm{obs}}), ~1\leq i \leq n_2.
  \end{equation}
  
  \subsubsection*{Mixing dynamic on latent arsenic surface}
  So far we have modeled $\theta_{1i}$ and $\theta_{2i}$: the static baseline log As surfaces at the same well $i$ in 2000--2001 and 2012--2013 respectively. Next, we model the temporal dynamic.  
  
  In an ideal noise-free isotropic  mixing scheme, the As concentration surface $y(x,t)$ as a function of location $x$ and time $t$ would follow a diffusion process,  
  $
  \frac{\partial }{\partial  t} y(x, t)  \propto \Delta_x  y(x, t).
  $
  Based on this heuristics,   we build an autoregression:
  \begin{equation}\label{eq_auto_reg}
  	\theta_{2i}=\theta_{1i}+\alpha_\delta+  (\beta_{\delta}+ \gamma_{i}) \delta_{i}  + \n(0,\tau), ~1\leq i \leq n_2.
  \end{equation}
  The temporal change between $\theta_{1i}$ and $\theta_{2i}$ contains the global shift $\alpha_\delta$,  the product of the mixing coefficient $\beta_{\delta}+ \gamma_{i}$ and the local Laplacian  $\delta_{i}$,  and independent   regression residuals.
  
  This autoregression dynamic \eqref{eq_auto_reg} is not equivalent to the diffusion process because it models $\theta$, the log scale As concentration. Besides, the discretization error over $t= 12$ years cannot be ignored. On the other hand, we do not expect the flow and recharge of groundwater As being perfectly described by a molecular diffusion and hence the diffusion process  itself is not realistic. 
  
  That being said, we view the  Laplacian  $\delta_{i}$ as an extracted feature that describes how the As concentration in a well is compared with its nearby neighbors. In order to learn the a more general  dynamic,  we allow the coefficient $\beta_{\delta}+ \gamma_{i}$ to vary by the initial well As, instead of a homogeneous diffusion constant.  In effect, we use \eqref{eq_auto_reg} to  learn a  mixing dynamic 
  \begin{equation}\label{eq_general_pde}
  	\frac{\partial}{\partial t} \theta (x, t) \propto \alpha_\delta+(\beta_{\delta}+ \gamma(\theta))\Delta_x (\theta).
  \end{equation} The regression residual in  \eqref{eq_auto_reg} determines the fidelity of the solution to this dynamic.
  
  \subsubsection*{Data-dependent mixing coefficient}
  The mixing coefficient 
  contains a constant term $\beta_{\delta}$ and a random term $\gamma_{i}$. We parameterize the latter term  by a function of $\theta_{1i}$,
  \begin{equation}
  	\gamma_{i}=\alpha_y \exp(\theta_{1i}/2) +  \alpha_\theta \theta_{1i}, 
  \end{equation}
  The combination of linear and exponential inputs is designed to fit both the large and small end of the initial value $\theta_{1i}$. 
  
  \subsubsection*{Priors and computation}
  Apart from the calibration parameter $\{c^\mathrm{cal}_k\}_{k=1}^9$, $\beta^{\mathrm{cal}}$ learned in advance,  the complete model contains 16950 free parameters in total: 
  $\{\beta_{l}\}_{0=1}^{485}$,   $\sigma_{\mathrm{obs}}$,  $\{\theta_{2i}\}_{i=1}^{8229}$, $\{\eta_{2i}\}_{i=1}^{8229}$,  $\alpha_0$, $\alpha_y$, $\alpha_\theta$,  $\alpha_\delta$, $\beta_\delta$, $\tau$, 
  on which we place weakly informative priors:
  \begin{equation}\label{eq_prior}
  	\begin{gathered}
  		\beta_{0}  \sim \n(4,2), ~ \beta_{l} \sim \n(0,0.5), 1\leq l \leq L=485.\\
  		\alpha_y\sim \n (0,0.2), ~ \alpha_\theta\sim \n (0,0.5), ~\alpha_\delta \sim \n (0,1), ~ \beta_\delta \sim \n (0,1).\\
  		\sigma_{\mathrm{obs}}\sim  \mathrm{InvGamma}(5,5), ~ \tau\sim \mathrm{InvGamma}(5,5).
  	\end{gathered}
  \end{equation}
  
  We fit the complete model \eqref{eq_log_y}--\eqref{eq_prior}  in Stan based on 4 chains and 2000 posterior simulation draws of all these parameters.
  Besides using the usual dynamic Hamiltonian Monte Carlo   sampler, we employ sparse matrix algebra when computing the log joint density, since 
  the basis functions $B_l(x^{\mathrm{N}}_{1i}, x^{\mathrm{E}}_{1i})$, $B_l(x^{\mathrm{N}}_{2i}, x^{\mathrm{E}}_{2i})$ and their Laplacians $\Delta B$ are sparse matrices.

  \section{Inference results for blanket surveys}\label{sec_inf_result}
  \subsection{Individual prediction}
\begin{figure}
  	\centering
  	\includegraphics[width=\textwidth]{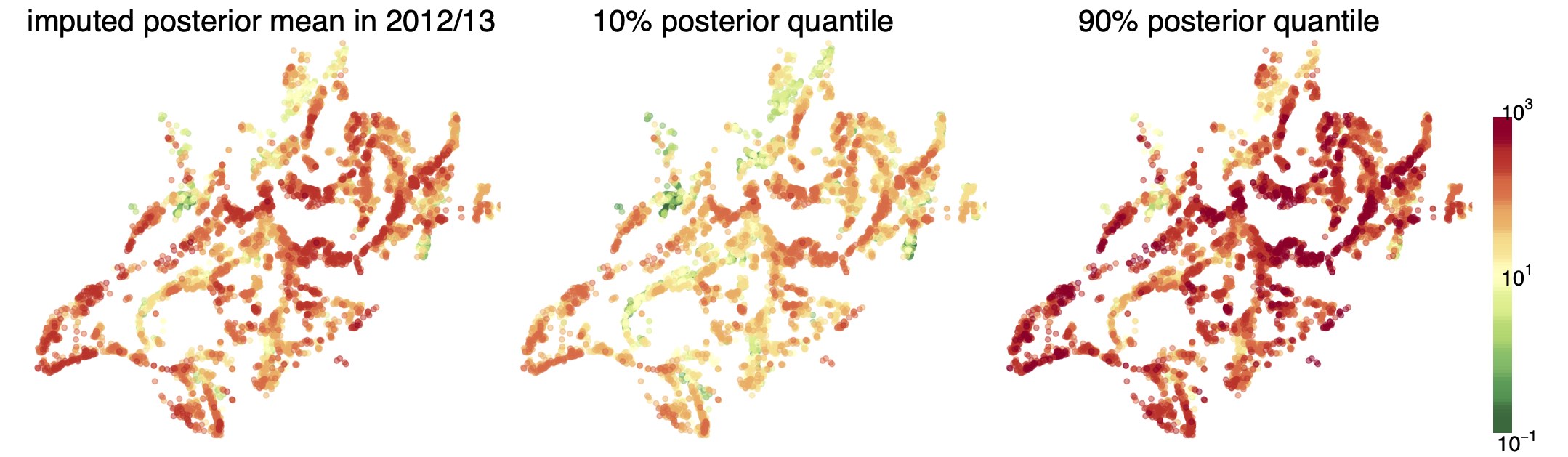}
  	\caption{The posterior mean, 10\%, and 90\% quantile of $\eta_2$: the calibrated  As level of the well in 2012--2013.}\label{fig_impute_value} 
  \end{figure}

\begin{figure}
\centering
  		\includegraphics[width=\textwidth]{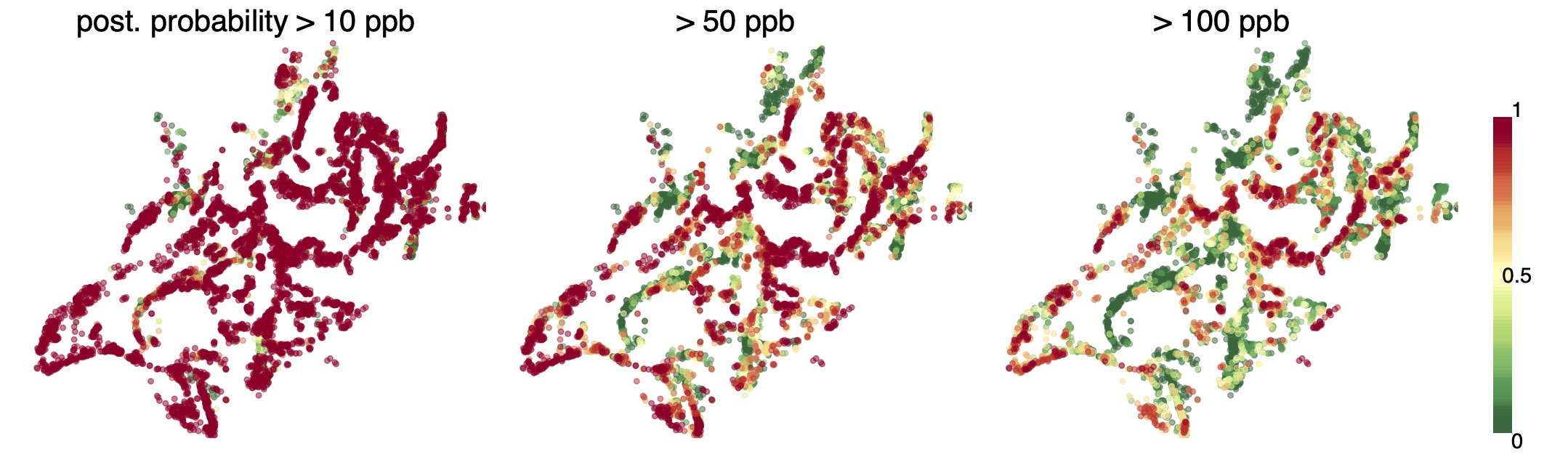}
  	\caption{The posterior probability that each sampled well was exceeding the As safety  threshold (10, 50, and 100 ppb) in 2012--2013.}  \label{fig_impute_prob} 
  \end{figure}

  The posterior distribution of $\eta_{2i}$ provides  inference for the As value in the $i$-th sampled well in 2012--2013. 
  It is both spatially smoothed owing to the spline, and  calibrated against the measurement errors.
  Figure \ref{fig_impute_value} displays the posterior mean and 10\% and 90\% quantile for all wells in 2012--2013. 
  For what matters to public health, we also compute the posterior probability of each well exceeding the safety threshold (10, 50, and  100 ppb) in Figure \ref{fig_impute_prob}.  

  \subsection{Mixing effect}
  \begin{figure} 
\centering
  		\includegraphics[width=\textwidth]{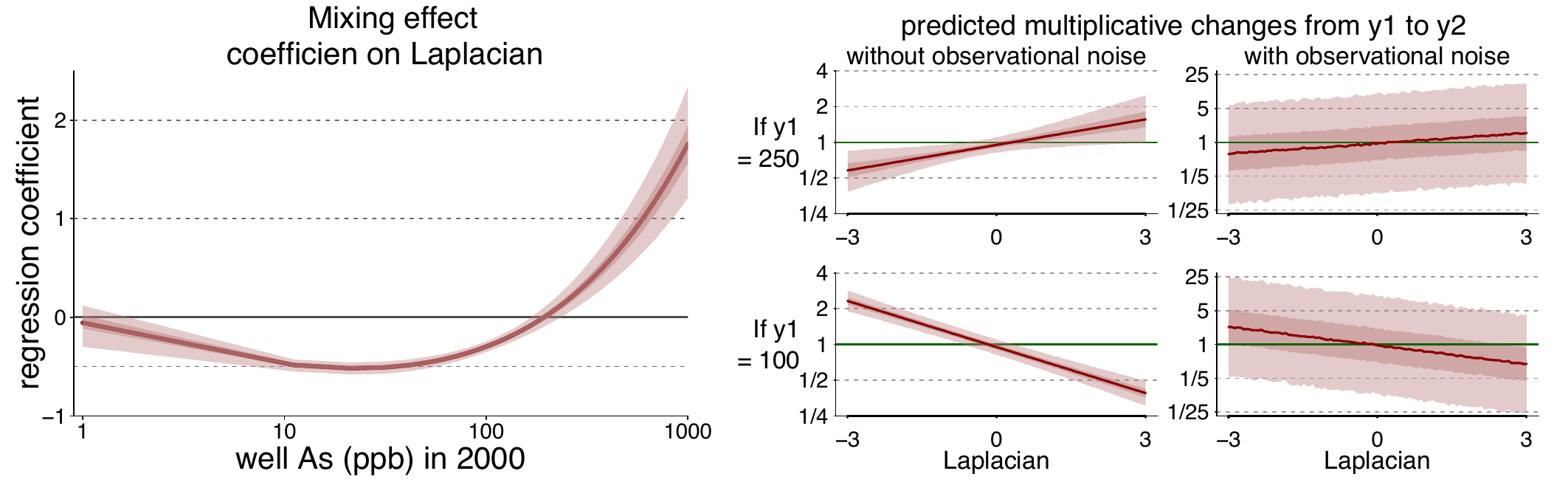} 
  	\caption{Left: the  posterior mean, 50\% and 95\% confidence interval of the regression coefficient before the Laplacian $\beta_{\delta}+ \alpha_y \exp(\theta_{1i}/2) +  \alpha_\theta \theta_{1i}$. 
  		A positive coefficient  represents a  diffusion effect, which is allowed to vary over $\theta_1$.   
  		Middle column: the  posterior mean, 50\% and 95\% confidence interval for predictive multiplicative change term $ \alpha_\delta+  (\beta_{\delta}+ \alpha (\theta_1)) \delta$ as a function of  $\delta$ for two fixed values  $\exp(\theta_{\theta_1})= 100$ and $250$ ppb.   Right:  posterior mean, 50\%, and 95\% confidence intervals for predictive multiplicative change in the observational level, with extra noise term added.}\label{fig_mixture_coef_log}
  \end{figure}

  \begin{figure}
  	\begin{center}
  		\includegraphics[width=\textwidth]{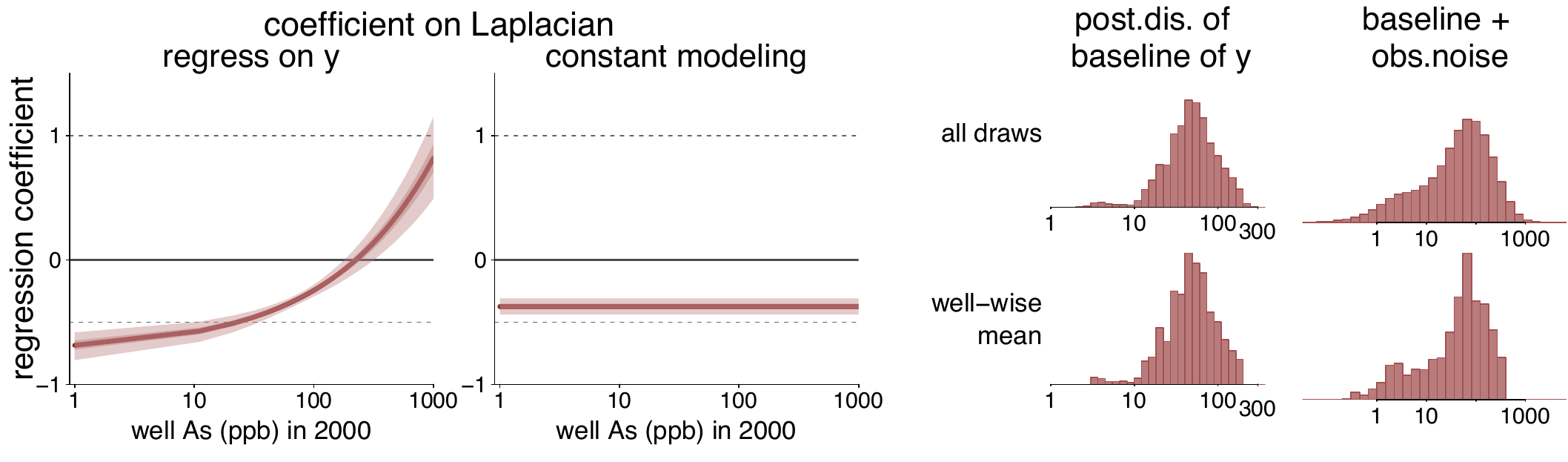} 
  	\end{center}
  	\caption{To check how sensitive the model is to the regression assumption, we consider two alternative models where the mixing effect coefficient is modeled by a a linear regression on $\exp(\theta)$ and a constant respectively. }\label{fig_mixture_coef_log_2}
  \end{figure} 
  
  The log As level changes from 2000--2001 to 2012--2013 in the $i$-th well is modeled by
  $\theta_{2i}-\theta_{1i} =   \alpha_\delta+  (\beta_{\delta}+ \gamma_{i}) \delta_{i}+$ noise. 
  Figure  \ref{fig_mixture_coef_log} shows the posterior mean, 50\% and 95\% confidence interval of the mixing coefficient, $(\beta_{\delta}+ \gamma_{i})$,  as a function of the initial baseline As concentration $\exp(\theta_{1i})$. When the initial   value  is very high ($>200$ ppb), the mixing coefficient is positive, which would drive locally large As values in 2000--2001   to drop in 2012--2013.  For small $\theta_1$, 
  the mixing  coefficient is estimated to be negative. As a result, the portions of the aquifers that had  As levels lower than than both their local neighbors and 200 ppb in 2000--2001 are  likely to become safer in average over time. 
  This pattern is consistent with the trend displayed for the subset of resampled wells in Figure \ref{fig_spline}.
  
  The middle column in Figure \ref{fig_mixture_coef_log} simulates the predictive multiplicative change term $ \exp\left(\alpha_\delta+  (\beta_{\delta}+ \alpha (\theta_1)) \delta\right)$ as a function of the Laplacian $\delta$ conditional on two fixed initial values  $\exp(\theta_{1})= 100$ and $250$ ppb respectively.  To give a sense of the additional observational noise, in the rightmost column, we simulate from the actual observable change by adding back the observational noise   $\exp\left(\alpha_\delta+  (\beta_{\delta}+ \alpha (\theta_1)) \delta + \n(0, \sigma_{\mathrm{obs}}) +\n(0, \tau)\right).$
  
  Should we worry about the the restriction that the coefficient $\beta_{\delta}+ \alpha_y \exp(\theta_{1}/2) +  \alpha_\theta \theta_{1}$  only has two degrees of freedom? First this is already an extension from the  
  constant-diffusion-coefficient model. We also consider two alternative models that replace replacing this functional form  by a linear regression on $\exp(\theta_{1}$ or a constant. The fitted result is shown in Figure \ref{fig_mixture_coef_log_2}. The newly fitted linear model is close to our main model fit except for the low end of the range in As concentrations. 
  
  Notably, $\theta_{1i}$ itself is a latent variable and we do not know its range a priori. The third and fourth column in Figure \ref{fig_mixture_coef_log_2} exhibit the histogram of all posterior simulation draws of the $\theta_{1i}$, the histogram of well-wise posterior mean $\E_{\mathrm{post}}(\theta_{1i})$, and the same two distribution for observed quantity $\eta_{2i}$. The majority of $\theta_{1i}$ is supported on the  $\log [10,300]$ interval. Consequently, the diffusion coefficient, as a function of  $\theta_1$  will rely on extrapolation beyond this support.  This explains why different parametric assumptions yield similar inference on this interval.  
  
  \subsection{Overall trend}

  \begin{figure}
  	\begin{center}
  		\includegraphics[width=\textwidth]{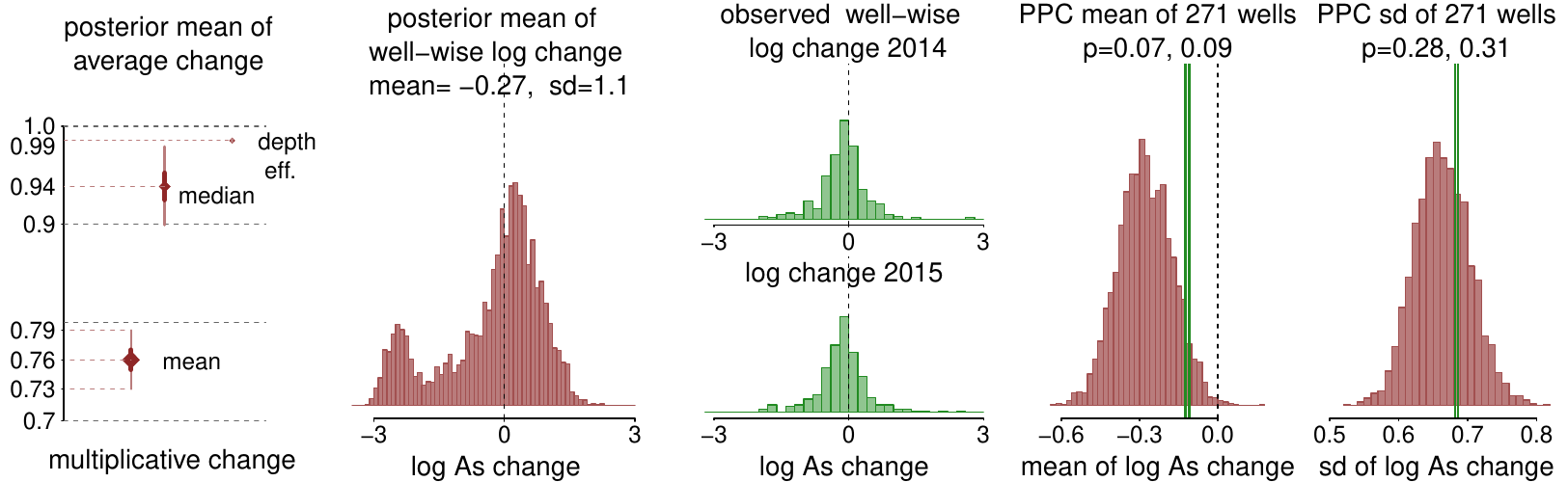} 
  	\end{center}
  	\caption{(1) Averaging over $n_2$ wells, the 
  		mean change from baseline $\theta_{1i}$ to $\theta_{2i}$ is equivalent  to a 76\% multiplicative change in As concentration, with 95\% confidence interval (73\%, 79\%). The median change is a $6\%$ decrease. Apart from the baseline $\theta$, the average depth of wells decreases 0.2 meter, adding to another 1\% drop.  
  		(2) The distribution of posterior mean of  $n_2$ individual well-wise log changes. 
  		(3) The observed well-wise change in the resampled 271 wells in 2014 and 2015. 
  		(4--5) The posterior predictive distributions of mean and standard deviation of log As change computed by 271 random wells each draw. They match with the observed mean and sd of log As changes in the resampled wells.} \label{fig_ppc}
  \end{figure}
  
  \begin{figure}[!ht]
  	\begin{center}
  		\includegraphics[width=\textwidth]{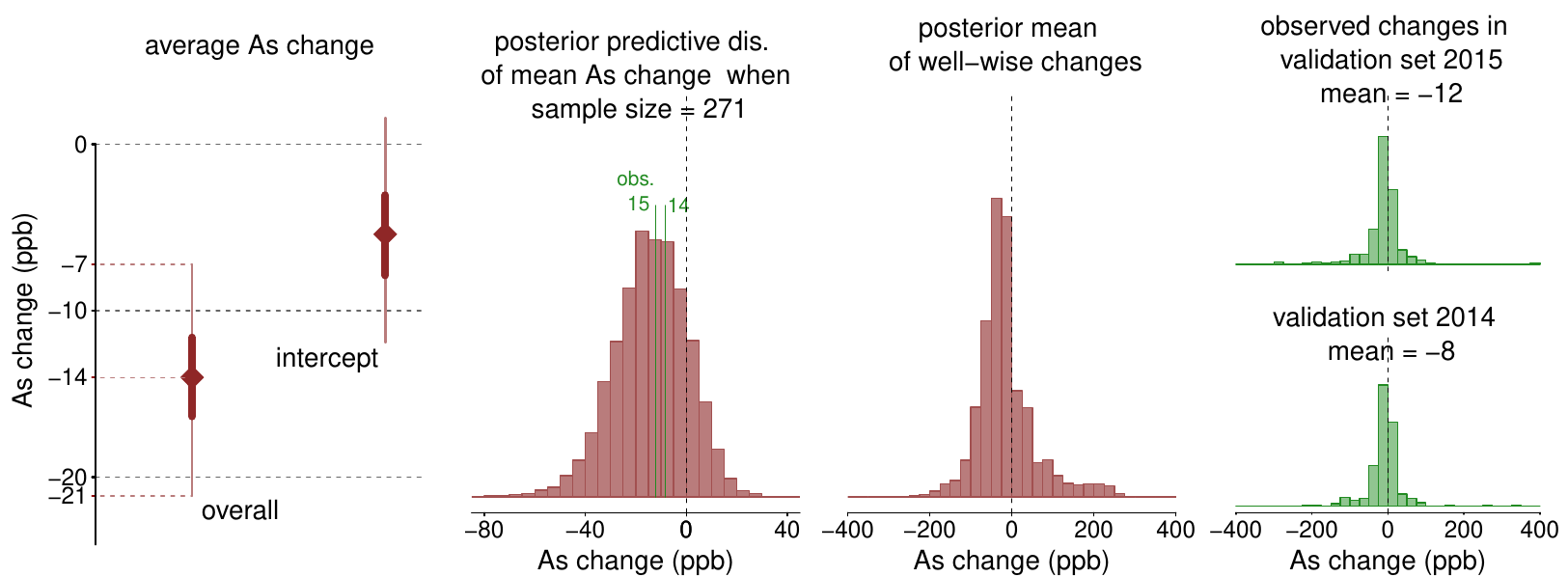}
  	\end{center}
  	\caption{(1) The posterior mean of overall As change averaged over all $n_2$ wells is 14 ppb decreasing, among which 5 ppb of the decrease is attributed to the intercept. (2) The posterior predictive distribution of overall As change averaged over 271 random wells. It matches with the observed mean As changes in the resampled wells.  (3) The distribution of posterior mean of  $n_2$ individual well-wise changes, among with 73\% of wells are inferred to decrease. (4) The observed well-wise change in the resampled wells.}   \label{fig_overall_change}
  \end{figure}
  Besides mixing effect, we are also interested in the direction and magnitude of  the  As levels change over time. In the left panel in Figure   \ref{fig_ppc},
  we simulate the average multiplicative well-wise  change among all wells in 2012--2013 by
  $ \exp(\frac{1}{n_2}\sum_{i=1}^{n_2} (\theta_{2i}-\theta_{1i}))$ and compute its  posterior mean and 95 \% confidence interval, which equals 76\% (95\% CI (73\%, 79\%)). In other words, the average As concentration (among all $n_2$ sampling locations) decreased by 24\% from 2000--2001 to 2012--2013.  
  Replacing the mean by median in the previous simulation, the median decrease is  6\% (95\% CI (2\%,10\%)).
  These inferred mean shifts  are evaluated at the same well in the same sampling location and is net of depth dependence. 
  The sample mean depth of 2012--2013 wells is slightly shallower (0.2 meter, or 1\%) than from 2000--2001, which contributes to an additional 1\% As mean decrease through the $\beta_\mathrm{depth} d$ term.
  
  
  However, a 24\% drop in average does not imply that concentrations in all wells had to decline over a decade. The second panel in Figure   \ref{fig_ppc} displays the histogram of posterior mean of well-wise  log change  $\theta_{2i}-  \theta_{1i}$: the log baseline change in the $i$-th well. The mean is $-0.27$ (equivalently 76\% in multiplicative scale) while the standard deviation of 1.1 is much larger. This large standard deviation reflects the intrinsic variation of individual level well-As changes, and will not shrink with a larger sample size. 
  
  To better interpret this temporal change,  we  draw posterior simulations of the  the linear scale As change (in ppb) of each well. The hypothetically-lab-measured average well As in the 2012--2013 blanket survey is   $\frac{1}{n_2} \sum_{i=1}^{n_2}\exp(\eta_{2i})$, which has a posterior mean of  $96$  ppb ($95$\% CI: $(94, 99)$).  The hypothetically-lab-measured well As at these locations  in 2000--2001 is  $1/n_2 \sum_{i=1}^{n_2}(\exp(\theta_{1i}+\beta_\mathrm{depth} (d_{2i}-d_0)+\n(0, \sigma_{\mathrm{obs}})))$, which has a posterior mean of $110$ ppb ($95$\% CI: $(104, 117)$). As for a reference, the sample mean of the first blanket survey in 2000--2001 is $109.6$ ppb. For each well in the second blanket survey, we can compute the paired change  $\exp(\eta_{2i})-\exp(\theta_{1i}+\beta_\mathrm{depth} (d_{2i}-d_0)+\n(0, \sigma_{\mathrm{obs}}))$.   Averaged over $i=1, \dots, n_2$,  the mean well As change is $-14$ ppb with $95$\% confidence interval $(-7, -21)$, as shown in the first panel in Figure \ref{fig_overall_change}. More precisely, the model  interprets  the overall log change $\theta_{2}-\theta_{1}$   into two terms, the intercept $\alpha_{\delta}$ and the diffusion effect $(\beta_{\delta}+\gamma_i)\delta_i$. The intercept represents the expected change if the mixing process akin to diffusion is completely blocked. We can generate the linear scale of posterior distribution of  the  intercept by simulation draws of $\frac{1}{n_2}\sum_{i=1}^{n_2} \exp(\eta_{2i}) (1- \exp(\alpha_{\delta}))$, which has a posterior mean of $-5$ ppb.  
  
  Again, a 14 ppb decrease in average As concentration does not mean all wells have to drop.  The third panel in Figure \ref {fig_overall_change} is the distribution of the well-wise posterior mean of As change, among which 27\% of individual wells are inferred to increase.

  \subsection{Posterior predictive model check on external validation data}\label{sec_vali}
  In order to verify the pattern we discovered in the blanket survey data,  we use the resampled 271 wells as a validation set to check the model fit of the blanket survey data.    As a caveat, the  resampled wells are not spatially identical to the blanket survey: the former one is relative coarse in the  northeast, such their sample average of lab measurements in 2000--2001 also differ slightly. Nevertheless, we still expect them to share a similar overall trend.    
  
  Because the well As is highly skewed in the blanket survey sample, the inferred mean log change ($-0.27$, or $-24$\% in multiplicative scale) is not the same as the log change of mean (the average well As drops from $110$ to $96$ ppb, or $-12\%$) at the same locations.  We check the model fit in both scales. 
  
  We first check the log changes of individual wells. We may compare the inferred  well-wise posterior mean of  log well As change among the blanket survey data (the second panel in Figure \ref{fig_ppc}) with the observed  well-wise log As changes in the resampled dataset  over a decade  (the third column in Figure \ref{fig_ppc}), but it is misleading as they differ in sample sizes. To make the correct model check,  for each joint simulation draw, we randomly choose 271 wells from the second blanket survey and compute the mean and standard deviation of the log As change among these 271 random wells.  The right two columns in Figure \ref{fig_ppc} visualize the posterior predictive distribution of these two quantities, both yielding acceptable $p$-values when compared with the observed validation data.

  We then the check the linear scale changes.  Again, we randomly draw a sample of size 271 among all $n_2$ wells in the second blanket survey and compute the sample average change. The second column of Figure  \ref{fig_overall_change} displays the histogram of 4000 posterior simulation draws of this random average, which matches well with the observed changes in both validation sets.

  \section{Discussion}\label{sec_discuss}
  \subsection{Limitations and assumptions}
  Our  model in section \ref{sec_reg} contains three  components: (a) imputing the counterfactual lab measurements from inaccurate kit data, (b) spatially smoothing the latent As surface,  and (c) estimating the mixing dynamic over time.

  Our approximation to the  differential equation \eqref{eq_general_pde} is in line with \citet{ramsay2007parameter}'s approach to parameter estimation in  general  differential equations by fitting well As surface using a collection of  basis functions. To this end,  the spatial modeling part \eqref{eq_bivariate_spline} describes the static log As surfaces by tensor products of $B$-spines. This spatial  component is similar to \citet{stone1988bivariate}'s approach to   approximating a thin plate spline. Instead of penalizing the roughness directly, the smoothness penalization in  our model is achieved by the  prior of spline coefficients \eqref{eq_prior}.
  The  $B$-spline approach was later criticized in spatial smoothing and interpolation for regions with  irregular shapes and boundaries, and its alternatives include products of $P$-splines \citep{eilers1996flexible},  soap-film splines \citep{wood2008soap},  and spatial spline regression \citep{sangalli2013spatial}. In our data (Figure \ref{raw_spatial_reidentified_wells}), the outer boundary of  the sampled wells is approximately convex, but there do exist several holes inside, because the wells are mainly located around the residential settlements, which are clustered and separated by open rice fields. Nevertheless, these human-settlement-holes do not define hard physical barriers for groundwater flows, hence we adopt simple Euclidean distance without extra modeling of the holes.

  Our analysis relies on several  assumptions. First, we modeled the log scale of the well As concentration and a constant observational error. We make this log transformation based on evidence from the residual plots in Figures \ref{temporal_change} and \ref{fig_cal}: After the log transformation of observations,  the error term has similar scales for different observations, which  otherwise is highly skewed in the linear scale. That said, we do not expect the observational error to be of exact same variance after the log transformation. This heterogeneity of  variance may still exist,  although it s partially remedied by  a flexible regression form (Gaussian process or bivaraite splines in our model). 
  Second, we learn the differentiation equation of the log As surface with one-step time-discretization (12 years). This is only reasonable  to explain the long term dynamic. We expect to better model the moderate the short term variation by through more frequent monitoring and  sampling in future research. 
  Third, the assumed dynamic \eqref{eq_general_pde} is isotropic, while the groundwater flow can be more complicated based on local geological structures. Fourth,  the $\theta_1$-varying mixing effect coefficient is only reasonable for the interval where $\theta_1$ is supported in observations. The claim on extreme values ($>$\SI{300}{\micro\gram\per\liter}) relies on extrapolations. Lastly, we model the dependence on well depth, but do not model other potential  selection mechanism of well-reinstallation, a potential confounder. Most households reinstalled their wells during the decade separating the two surveys but those who did so with the goal of reducing their arsenic exposure did so outside the depth range disturbed by irrigation pumping that was considered for this analysis \citep{jamil2019effectiveness}. The shallower well reinstallations analyzed here are therefore less likely to be biased.

  \subsection{Practical implications of inferred spatiotemporal trends}
  Our analysis shows considerable variability  in arsenic concentrations, but fortunately, no indication from the two larger surveys conducted a decade apart that low-arsenic portions of the shallow aquifer became  systematically contaminated over time. In particular, for wells meeting the local safety range (As $<$\SI{50}{\micro\gram\per\liter}), the inferred mixing dynamic suggests they are unlikely to be elevated due to neighboring influence.   Consequently, the  well-switching should be more systematically encouraged in Araihazar and many other parts of Bangladesh.  
  
  This mixing effect could have been a concern because groundwater flow patterns have been thoroughly altered by irrigation pumping drawing large volumes of water from shallower aquifers during each winter season throughout the region \citep{harvey2006groundwater}. 

  The documented decline in mean arsenic for the study region is of considerable interest as well because it provides new evidence that accelerated groundwater pumping has been withdrawing arsenic from the shallow aquifer, albeit at the cost of redepositing this arsenic in rice fields  \citep{dittmar2007spatial}. These withdrawals are compensated each year at the onset of the monsoon by recharge of low-arsenic surface water. The resulting flushing of the shallow aquifer has evidently only partially been compensated by a release of arsenic from arsenic sands through a previously proposed exchange process \citep{van2008flushing, Mozumder2019Impacts}.  The much larger pool of arsenic present in aquifer sediments compared to the pool of arsenic in groundwater has therefore delayed the decline in groundwater arsenic concentrations due to flushing but has probably also anchored the distribution of arsenic to the local geology. This has prevented greater convergence of concentrations to the areal mean due to the turning on and off of irrigation pumps. Such conclusions could not have been reached with a detailed statistical analysis of the available data.

  \subsection{The merit of field kits}
  In agreement with  with previous findings \citep{van2003spatial, mailloux2020recommended},  our model confirms a considerable variation in individual well As, both spatially (between neighboring wells or villages, explained by $\sigma_{\mathrm{obs}}$ in \eqref{eq_log_y}), and temporally  (despite an overall trend, individual wells are likely to drop or incline over a decade, explained by $\tau$ in \eqref{eq_auto_reg}). 
  
  The field kit measures As concentrations  with much less precision compared to laboratory measurements, exhibiting both large bias and variance.   On one hand, this measurement error amplifies the already-large noise-to-signal ratio in observations, on top of the spatial and temporal variations. Takings the face value of  field kit measurements increases the chance of incorrect and inconsistent labeling.
  That said,  laboratory precision is not necessary 
  when concentrations range across several orders of magnitude \citep{van2005reliability}.
  Health-based thresholds are often somewhat arbitrary. The health impacts of drinking water that contains 9 or \SI{11}{\micro\gram\per\liter} As are to first approximation a linear function of exposure and therefore not that different, for instance. What is  more important is for rural residents in Bangladesh to know if their well As is closer to 1 or  \SI{100}{\micro\gram\per\liter}.

  On the other hand,  field kits comes with the distinct advantage that the cost of the measurement is an order of magnitude lower and, perhaps, more importantly that the result can be delivered on the spot. Bringing a sample to the laboratory and the result back to the well owner is logistically much more complicated. By  incorporating  information from the quality-control sample, spatial smoothing, and dynamic pattern over time, our statistical model calibrates the individual filed kit results and makes personalized probabilistic prediction for each well (Figures \ref{fig_impute_value} and \ref{fig_impute_prob}), which is helpful for residents to distinguish current safe and unsafe wells, and instructive for determining future well-installation locations.    
  
  Looking forward, either for the purpose of  better understanding the dynamic pattern or monitoring individual household exposures, the high variability in groundwater As or  other toxicants in the environment  calls for more frequent and extensive sampling. Imprecise but widely-accessible field kit tests in companion with flexible statistical modeling that facilitates this open-ended data gathering can provide a balance between total cost and accuracy in many areas of geoscience research and policy.

 \section*{Acknowledgements}
 Data collection in the field and in the laboratory was supported in part by NIEHS Superfund Research Program grant P42 ES010349 and
 NSF grant ICER 1414131.  We also thank  the National Science Foundation, Institute of Education Sciences, Office of Naval Research, National Institutes of Health, Sloan Foundation, and Schmidt Futures for financial support and Daniel Simpson for helpful discussions.

 \vspace{1cm}
 \bibliographystyle{apalike} 
 \bibliography{arsenic}	
 \setlength{\bibsep}{0.9pt}
 
 \newpage
 \appendix
 \section{Supplementary material}
\subsection{Replication data and code}  Available at { \url{https://github.com/yao-yl/As-measurement-code}}.

\subsection{Graphical representation of models}
\begin{figure}[!ht]
	\centering
	\begin{tikzpicture}
		\tikzstyle{main}=[circle, minimum size = 3mm, thick, draw =black!80, node distance = 14mm]
		\tikzstyle{connect}=[-latex, thick]
		\tikzstyle{box}=[rectangle, draw=black!100]
		\node[main, fill = white!100] (theta1) [label=above:$\theta_{2000}$  ] { };
		\node[main] (theta2) [ right=40 mm   of theta1,label=above:$\theta_{2014} \quad  \approx$] {};
		\node[main] (theta3) [ right= 10 mm    of theta2,label=above:$\theta_{2015}$] { };
		\node[main, fill = gray] (y1) [below=5 mm   of theta1,  label=below:$\tilde y_{2000}$] { };
		\node[main, fill = gray] (y2) [ right=40 mm   of y1,label=below:$\tilde y_{2014}$] {};
		\node[main, fill = gray] (y3) [ right= 10 mm    of y2,label=below:$\tilde y_{2015}$] { };
		\path   (theta2) edge [connect] (y2)
		(theta1) edge [connect] (y1)
		(theta1) edge [connect] (theta2)
		(theta2) edge [connect] (theta3)
		(theta3) edge [connect] (y3);
		\node[rectangle, inner sep=3.5mm, fit= (theta1) ,  label= left:  \emph{process model}, xshift=-2mm] {};
		\node[rectangle, inner sep=3.5mm, fit= (y1) ,  label= left:  \emph{observation model}, xshift=-2mm] {};
		\node[rectangle, inner sep=3.5mm, fit= (theta1) ,  label= above: {\em Gaussian process prior}] {};
		\node[rectangle, inner sep=5mm, fit= (theta1) ,  label= right: {\em cubic spline},  yshift=2mm  ] {};
	\end{tikzpicture}\caption{Graphical summary of  the  model for  271 resampled wells. The observed arsenic concentration $\tilde y$ is modeled by a noisy realization of  the baseline value $\theta$, and  $\theta$ is placed a Gaussian process prior for spatial smoothness. The change from  $\theta_{2010}$ to  $\theta_{2014}$ and $\theta_{2015}$ is modeled by an auto regression  with cubic splines. }\label{fig_model}
\end{figure}
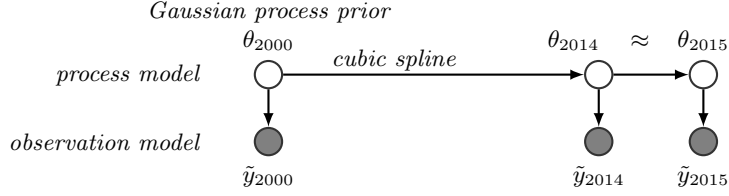

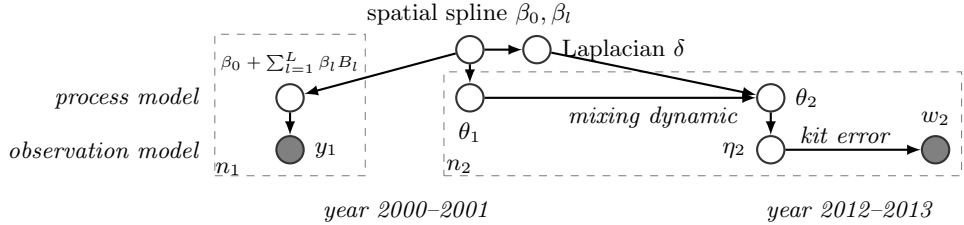
\begin{figure}[!ht]
	\centering
	\begin{tikzpicture}
		\tikzstyle{main}=[circle, minimum size = 2mm, thick, draw =black!80, node distance = 14mm]
		\tikzstyle{connect}=[-latex, thick]
		\tikzstyle{box}=[rectangle, draw=black!100]
		\node[main, fill = white!100] (theta1) [label=above:{\tiny $\beta_{0}+ \sum_{l=1}^L\beta_{l} B_l$} ]{ };
		\node[main] (theta2) [ right=60 mm   of theta1,label=right:$\theta_{2}$] {};
		\node[main] (theta21) [ right =20 mm   of theta1,label=below:$\theta_{1}$] {};
		\node[main, fill = gray] (y1) [below=3 mm   of theta1,  label=right:$y_{1}$] { };
		\node[main] (beta) [above=2.5 mm   of theta21,  label=above:{ spatial spline $\beta_0, \beta_{l}$}] { };
		\node[main] (delta) [right =5 mm   of beta,label=right: {Laplacian $\delta$}] {};
		\node[main] (y2) [ right=60 mm   of y1,label=left:$\eta_{2}$] {};
		\node[main, fill = gray] (y22) [ right=18 mm   of y2 ,label=above:$w_{2}$] {};
		\node[rectangle, inner sep=.5mm, fit= (y2) ,   label= right: \emph{kit error}, xshift=0mm, yshift=2mm] {};
		\node[rectangle, inner sep=3.5mm, fit= (theta1) ,  label= left: \emph{process model}, xshift=-5mm] {};
		\node[rectangle, inner sep=3.5mm, fit= (y1) ,   label= left:  \emph{observation model}, xshift=-5mm] {};
		\node[rectangle, inner sep=3.5mm, fit= (y1) ,   label= below:  \emph{ year 2000--2001}, xshift=15mm] {};
		\node[rectangle, inner sep=3.5mm, fit= (theta21) ,   label= right:  \emph{ mixing dynamic}, xshift=5mm, yshift=-2.5mm] {};
		\node[rectangle, inner sep=3.5mm, fit= (y2) ,   label= below:  \emph{ year 2012--2013}, xshift=10mm] {};
		\node[rectangle, inner sep=1.5mm, fit= (y22)(theta21),    xshift=0mm, draw=gray, dashed](plate2) {};
		\node[rectangle, inner sep=4mm, outer sep=3pt, fit= (y1)(theta1) ,   yshift=2.5mm, minimum width=2cm, draw=gray, dashed](plate1) {};
		\node[anchor=south west,inner sep=3pt] at (plate1.south west) 
		{\footnotesize $ n_1$};
		\node[anchor=south west,inner sep=1pt] at (plate2.south west) 
		{\footnotesize $ n_2$};
		\path   (theta2) edge [connect] (y2)
		(theta1) edge [connect] (y1)
		(y2) edge [connect] (y22)
		(theta21) edge [connect] (theta2)
		(beta) edge [connect] (theta1)
		(beta) edge [connect] (theta21)
		(beta) edge [connect] (delta)
		(delta) edge [connect] (theta2);
	\end{tikzpicture}\caption{  Graphical summary of the model for blanket survey data (section \ref{sec_reg}). We model the spatial distribution by a spline, and the temporal  dynamic is modeled by an autoregression on the initial baseline value and the  Laplacian.  The field kit is calibrated back to the hypothetical lab reading $\eta_2$, and further linked to the baseline value $\theta_2$ via a Gaussian noise.  }\label{fig_graph2}
\end{figure}

Figure \ref{fig_model} and \ref{fig_graph2} summarize the model for 271 re-identified wells (section \ref{sec_model_subset}), and the model for blanket survey data (section \ref{sec_reg}, the main model) in graphs. 

\subsection{Knots of bivariate \texorpdfstring{$B$}{B}-spline}\label{sec_spline}
\begin{figure}
	\centering
	\includegraphics[width=\textwidth]{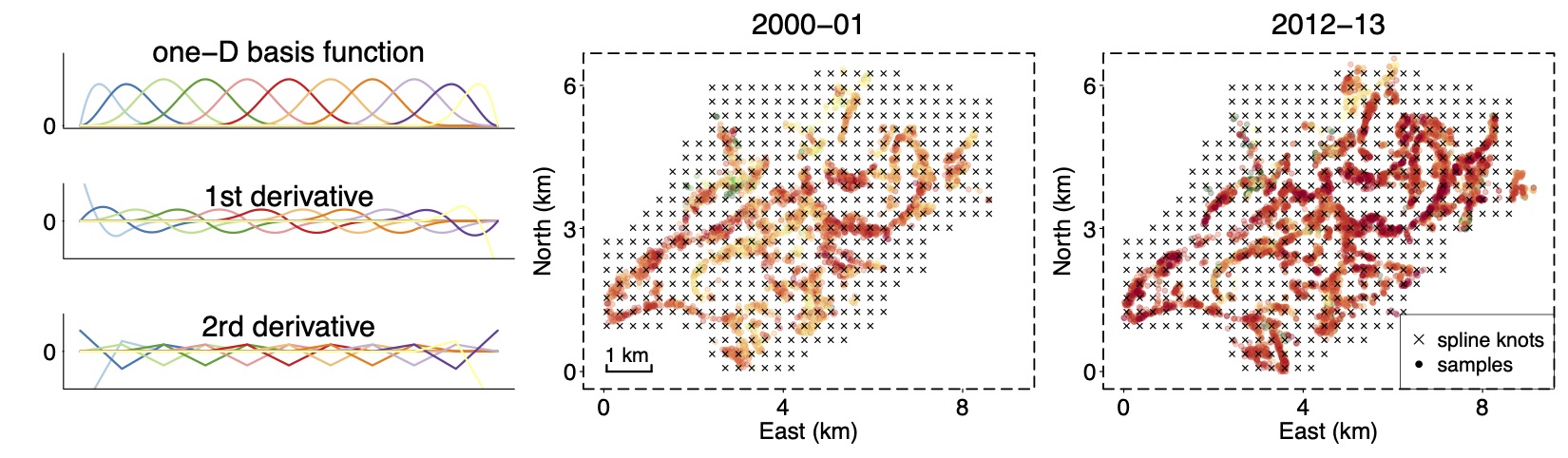}
	\caption{ Left: Illustration of one-dimensional cubic spline basis functions and their first two derivatives. Right two panels: We build the bivariate splines from the tensor product of one dimensional $B$-splines. The knots are represented by black crosses on the graph. They are  equally-spaced,  and trimmed outside sample range.} \label{spline_basis}
\end{figure}
In section \ref{sec_reg}, the spatial modeling \eqref{eq_log_y} relies on bivariate $B$-splines. 
We place the inner knots on a uniform  grid spanned by equally-spaced vertical coordinates 
$\tilde x^{\mathrm{N}}_1, \dots, \tilde x^{\mathrm{N}}_{N_1}$ and horizontal coordinates  $\tilde x^{\mathrm{E}}_1, \dots, \tilde x^{\mathrm{E}}_{N_2}$ such that  $\tilde x^{\mathrm{N}}_{i+1}-\tilde x^{\mathrm{N}}_{i}= \tilde x^{\mathrm{E}}_{j+1}-\tilde x^{\mathrm{E}}_{j}=x_0$. In other words, the inner knots are placed on the vertexes of $x_0 \times x_0$ blocks.

The  spacing is chosen such that there are 30 inner knots of longitudes, which results in $x_0=293$ meters, $N_1=30$, and $N_2=22$. For a comparison, the posterior mean of the  length scale $\rho$ from  the Gaussian process regression \eqref{model_gp} is estimated to be 381 meters.

On each coordinate, we construct the usual cubic spline basis functions with knots $\tilde x^{\mathrm{N}}_1, \dots, \tilde x^{\mathrm{N}}_{N_1}$ or   $\tilde x^{\mathrm{E}}_1, \dots, \tilde x^{\mathrm{E}}_{N_2}$.  The number of basis functions on each dimension is $N_1+ 3 = 33$ and $N_2 +3 = 25$.
We consider the tensor product of these two dimensions, 
$$ B_l(x^{\mathrm{N}}_{2i}, x^{\mathrm{E}}_{2i})  =  B^\mathrm{N}_l(x^{\mathrm{N}}_{2i})  B^\mathrm{E}_l(x^{\mathrm{E}}_{2i}),$$
where $l$ is the rearranged index of the product basis functions, ranging from 1 to $25\times 33=825$.  

We prune the knots by remove the blocks not covering any observed wells from blanket surveys. The pruned knots have holes inside, and we remove these holes by completing the convex hull of the pruned knots.  After this procedure, the number of  not-constant-zero basis function is $L=485$. We relabel the index $l$ from 1 to 485. We plot the trimmed knots with reference to observations in Figure \ref{spline_basis}. 

The basis function is evaluated at location $x_{1i}$ and  $x_{2i}$, which are stored by two matrices $B(x_1)$ and  $B(x_2)$  with dimension $4574\times  485$ and $8295 \times 485$ respectively.       

Finally we compute the Laplacian of basis functions. This is straightforward for tensor products for 
$$   \Delta  B_l(x^{\mathrm{N}}_{2i}, x^{\mathrm{E}}_{2i})  = \Delta  B^\mathrm{N}_l(x^{\mathrm{N}}_{2i})  B^\mathrm{E}_l(x^{\mathrm{E}}_{2i})   + \Delta  B^\mathrm{E}_l(x^{\mathrm{E}}_{2i}) B^\mathrm{N}_l(x^{\mathrm{N}}_{2i}).$$

We store this product  $\Delta B(x_2)$, Laplacian basis evaluated at $x_2$,     by  a $8295 \times 485$ matrix. 

The matrix of  basis functions
$B(x_1)$, $B(x_2)$ and $\Delta B(x_2)$ only  come into the log joint density via matrix multiplication, during which  employ sparse matrix algebra.

\subsection{Imputation result of pointwise Laplacian}

\begin{figure}
	\centering
	\includegraphics[width=0.5\textwidth]{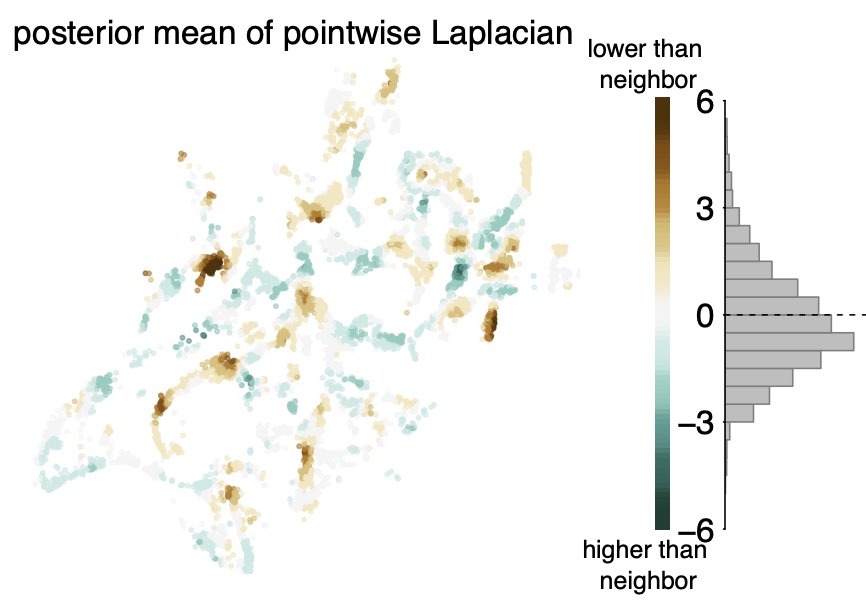}
	\caption{The posterior mean of the Laplacian $\delta_i$ in year 2000. A positive Laplacian means the well has lower As value than its neighboring wells. On the right is the histogram of the posterior mean of well-wise Laplacian.}\label{fig_impute_lap} 
\end{figure}
\begin{figure}[!ht]
	\centering
	\includegraphics[width=\textwidth]{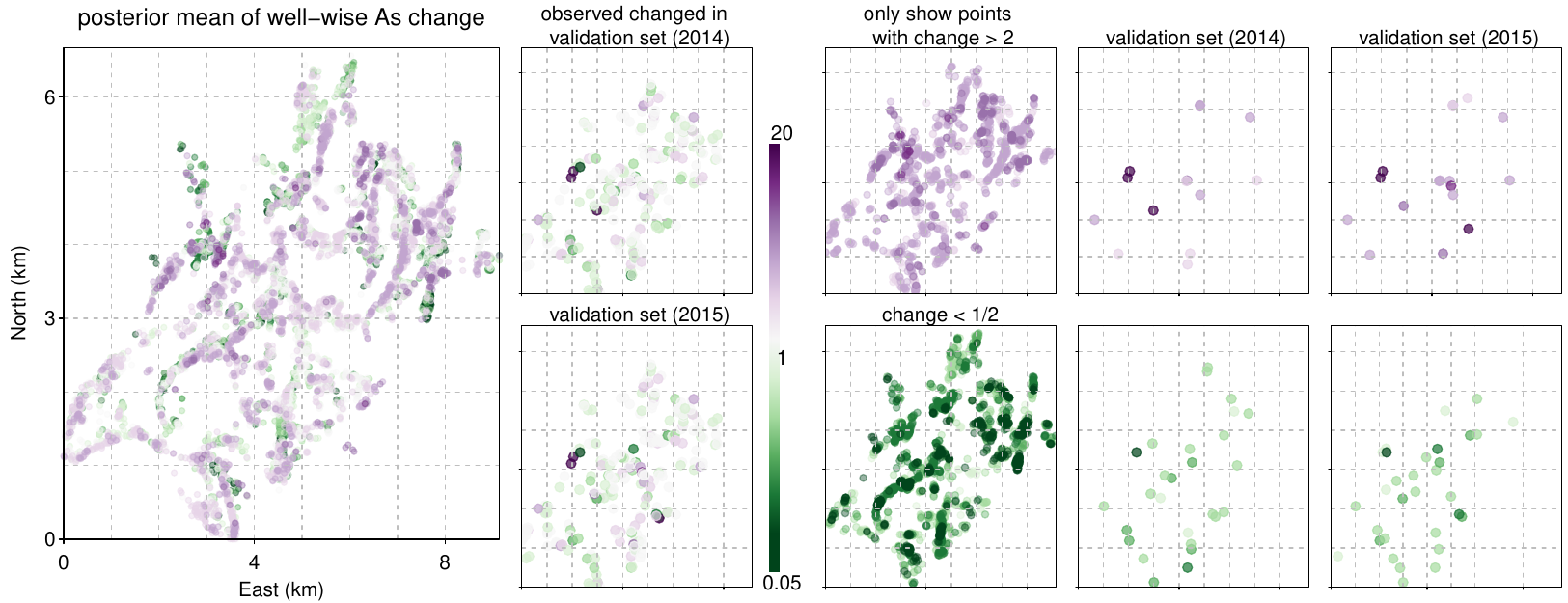}
	\caption{Column (1) The posterior mean of inferred log As changes of $n_2$ individual well: $\theta_{2i}-\theta_{1i}$. (2) The observed log changes in the 271 resampled well. (3-5) The subset of  wells with inferred or observed changes more than doubling (first row) or less than halving (second row). The fitted model is able to match certain local patterns in the validation set.}\label{fig_spatio_change}  
\end{figure}
Figure \ref{fig_impute_lap} displays the posterior mean of the Laplacian  $\delta_i$ from the fitted blanket survey data. A positive Laplacian means the well has lower As value than its neighboring wells.  
The Laplacian is not scaling invariant. In our computation, we standardize the input matrix $\Delta B$ by a factor 1000, and standardize the location coordinate $x^{\mathrm{N}}, x^{\mathrm{E}}$ by a factor 9.1 km (so as to make them unit-scaled).
To interpret the scale of the Laplacian, we use the heuristic from the finite difference approximation 
\begin{align*}
	\Delta\theta\left( x_{2}^{\mathrm{N}}, x_{2}^{\mathrm{E}}\right)\approx & \frac{1}{4}h^2\bigl( \theta\left( x_{2}^{\mathrm{N}}+h, x_{2}^{\mathrm{E}}\right) + \theta\left( x_{2}^{\mathrm{N}}-h, x_{2}^{\mathrm{E}}\right) \\
	&+ \theta\left( x_{2}^{\mathrm{N}}, x_{2}^{\mathrm{E}}+h\right) + \theta\left( x_{2}^{\mathrm{N}}, x_{2}^{\mathrm{E}}-h\right)- 4\theta\left( x_{2}^{\mathrm{N}}, x_{2}^{\mathrm{E}}\right)\bigr).    
\end{align*}
Using the average  closest distance in the second survey, $h=28$ meters, a unit scale of Laplacian  is approximately  $\exp(1000(28/9100)^2)= 1.01$ in multiplicative factor, i.e., Laplacian= $\pm 1$ equivalents being 1\% lower/higher than all closest well in the neighborhood.

One caveat is that the pointwise value of the Laplacian $\delta_i$ is unlikely to be estimated accurately,  not only because the latent variable $\theta$ is not observable, but also because the spline fit implicitly penalizes the roughness. That said, we view the fitted $\delta_i$ as an extracted  feature to  represent the  neighboring difference. 


\emph{Posterior predictive check of the inferred Laplacian}. How reliable is our conclusion concerning the diffusion-like process of homogenisation of arsenic concentrations? The challenge in checking the model is that all associated quantities describing the process, the baseline log As value $\theta_{1}, \theta_{2}$ and the pointwise Laplacian, are latent variables. 
Nevertheless, we have already seen a consistent pattern in Figure \ref{fig_spline} and \ref{fig_mixture_coef_log}. We further check the model claim by visually checking if the model is able to pick the spatially local change pattern. The leftmost panel in Figure \ref{fig_spatio_change} visualizes the posterior mean of inferred log As changes of $n_2$ individual well: $\theta_{2i}-\theta_{1i}$, and the second column is the observed log changes in the 271 resampled well. To reduce overlapping, we also draw the subset of wells that have inferred or observed changes more than doubling or less than halving. Although we cannot completely separate noise, the fitted model is able to match certain local patterns in the validation set. The local regions with drastic changes 
are colored in agreement in the fitted model.

\subsection{Dependence on well depth}
\begin{figure}[!hb]
	\begin{center}
		\includegraphics[width=\textwidth]{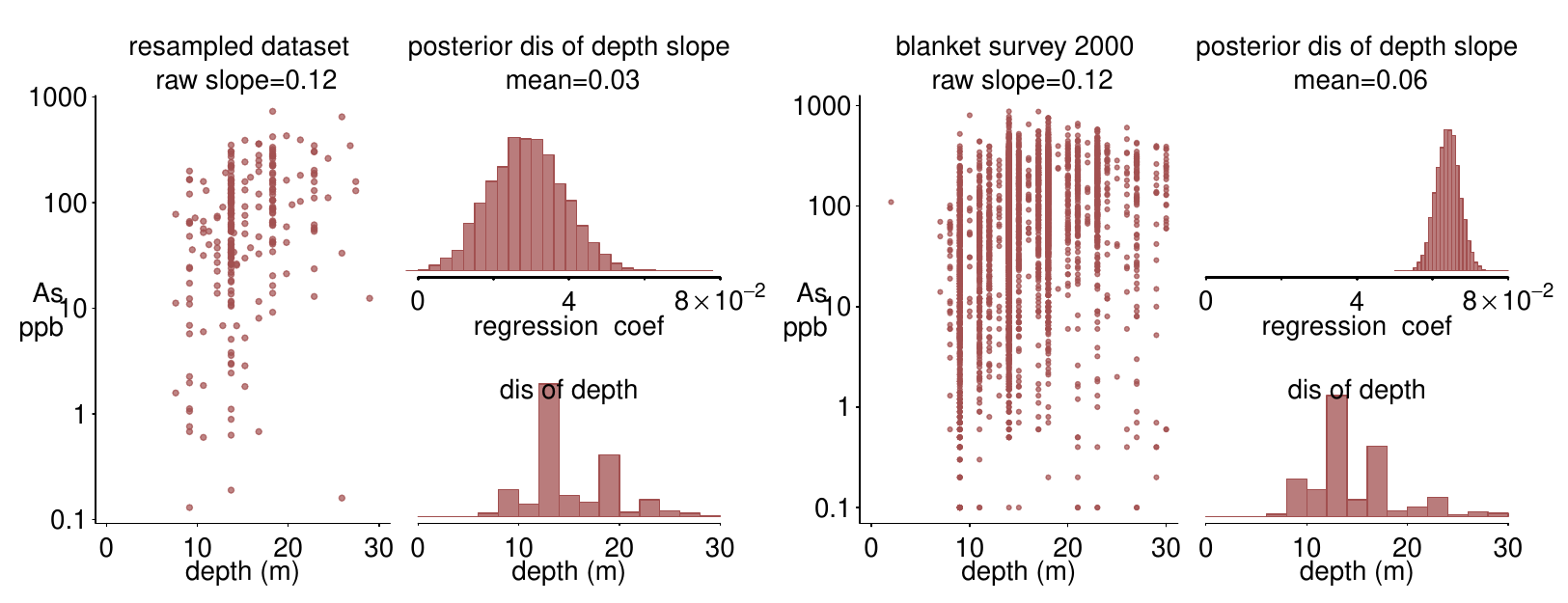}
		\caption{Dependence of well-As on well depth.
			First column: the observed As concentration-well depth in the 271 resampled dataset in 2015. Second column: the posterior distribution of $\beta_\mathrm{depth}$ from the first model and resampled dataset, and the histogram of the  well depth in the sample.  Right two columns: the same scatterplot and histogram, inferred using the second model and larger dataset.}\label{fig:depth}
	\end{center}
\end{figure}

The first and third panel in Figure \ref{fig:depth} displays the relation between well depth and log As concentration in the resampled dataset and blanket survey 2000. A direct univariate regression has coefficient 0.12, which ignores the spatial dependence on well depth. In our model, the posterior mean of $\beta_\mathrm{depth}$ is 0.03 and 0.064 (95\% CI: (0.057, 0.07)) in the resampled and blanket-survey dataset. The latter one has smaller standard deviation for its larger sample size.   

It is previously known that among  shallow ($<$30 m) wells, the well As is positively correlated with well depth even after controlling the spatial variation \citep{gelman2004direct}, although  this correlation does not necessarily imply a causal relation between them.

Our model for blanket surveys contains both the spatial distribution and the depth dependence. The well-wise are therefore the controlled-comparison: the inferred well-wise change given the same well depth.  Overall the depth dependence is weak compared with spatiotemporal variations. From 2000--2001 to 2012--2013, the average well depth decreased by 0.23 meters, which adds to $1-\exp(-0.2\times0.06)=1$\% decrease in the sample mean As concentration.

\end{document}